\documentclass[twocolumn]{aastex631}
\usepackage{multirow}

\newcommand{\sups}[1]{$^{#1}$}

\newcommand{\hbeta}{H$\beta$}

\newcommand{\oiiir}{[O\thinspace{\sc iii}]$\lambda5008$}
\newcommand{\oiii}{[O\thinspace{\sc iii}]$\lambda\lambda4960,5008$}
\newcommand{\oiiix}{[O\thinspace{\sc iii}]}

\newcommand{\halpha}{H$\alpha$}
\newcommand{\niir}{[N\thinspace{\sc ii}]$\lambda6585$}
\newcommand{\nii}{[N\thinspace{\sc ii}]$\lambda\lambda6548,6585$}
\newcommand{\niix}{[N\thinspace{\sc ii}]}
\newcommand{\oiiihb}{\oiiir/\hbeta}
\newcommand{\niiha}{\niir/\halpha}

\newcommand{\Rsig}[1]{R{#1}$\sigma$}

\begin{document}

\title{MOSDEF-3D: Keck/OSIRIS Maps of the Ionized ISM in $\bf{z\sim2}$ Galaxies}

\author[0000-0001-9489-3791]{Natalie Lam}
\altaffiliation{NSF Graduate Research Fellow}
\affiliation{Department of Physics \& Astronomy, University of California, Los Angeles, 430 Portola Plaza, Los Angeles, CA 90095, USA}

\author[0000-0003-3509-4855]{Alice E. Shapley}
\affiliation{Department of Physics \& Astronomy, University of California, Los Angeles, 430 Portola Plaza, Los Angeles, CA 90095, USA}

\author[0000-0003-4792-9119]{Ryan L. Sanders}
\affiliation{Department of Physics and Astronomy, University of Kentucky, 505 Rose Street, Lexington, KY 40506, USA}

\author[0000-0001-9554-6062]{Tuan Do}
\affiliation{Department of Physics \& Astronomy, University of California, Los Angeles, 430 Portola Plaza, Los Angeles, CA 90095, USA}

\author[0000-0001-5860-3419]{Tucker Jones}
\affiliation{Department of Physics and Astronomy, University of California Davis, 1 Shields Avenue, Davis, CA 95616, USA}

\author[0000-0002-2583-5894]{Alison Coil}
\affiliation{Department of Astronomy \& Astrophysics, University of California, San Diego, La Jolla, CA 92092, USA}

\author[0000-0002-7613-9872]{Mariska Kriek}
\affiliation{Leiden Observatory, Leiden University, NL-2300 RA Leiden, Netherlands}

\author[0000-0001-5846-4404]{Bahram Mobasher}
\affiliation{Department of Physics \& Astronomy, University of California, Riverside, 900 University Avenue, Riverside, CA 92521, USA}

\author[0000-0001-9687-4973]{Naveen A. Reddy}
\affiliation{Department of Physics \& Astronomy, University of California, Riverside, 900 University Avenue, Riverside, CA 92521, USA}

\author[0000-0002-4935-9511]{Brian Siana}
\affiliation{Department of Physics \& Astronomy, University of California, Riverside, 900 University Avenue, Riverside, CA 92521, USA}

\author[0000-0003-1249-6392]{Leonardo Clarke}
\affiliation{Department of Physics \& Astronomy, University of California, Los Angeles, 430 Portola Plaza, Los Angeles, CA 90095, USA}

\begin{abstract}

We present spatially-resolved rest-frame optical emission line maps of four galaxies at $z\sim2$ observed with Keck/OSIRIS to study the physical conditions of the ISM at Cosmic Noon. Our analysis of strong emission line ratios in these galaxies reveals an offset from the local star-forming locus on the BPT diagram, but agrees with other star-forming galaxies at similar redshifts. Despite the offset towards higher \oiiihb{} and \niiha{}, these strong-line ratios remain consistent with or below the maximum starburst threshold even in the inner $\sim1$~kpc region of the galaxies, providing no compelling evidence for central AGN activity. The galaxies also exhibit flat radial gas-phase metallicity gradients, consistent with previous studies of $z\sim2$ galaxies and suggesting efficient radial mixing possibly driven by strong outflows from intense star formation. Overall, our results reveal the highly star-forming nature of these galaxies, with the potential to launch outflows that flatten metallicity gradients through significant radial gas mixing. Future observations with JWST/NIRSpec are crucial to detect fainter emission lines at higher spatial resolution to further constrain the physical processes and ionization mechanisms that shape the ISM during Cosmic Noon.
\end{abstract}


%%%%%%%%%%%%%%%%%%%%%%%%%%%%%%%%%%%%%%%%%%%%%%%%%%%%%%%%%%%%%%%%%%%%%%%%%%%%%%%%%%%%%%%%%%%%%%%%%%
%%%%%%%%%%%%%%%%%%%%%%%%%%%%%%%%%%%%%%%%%%%%%%%%%%%%%%%%%%%%%%%%%%%%%%%%%%%%%%%%%%%%%%%%%%%%%%%%%%
\section{Introduction} \label{sec:intro}
 
During the peak of cosmic star formation activity, commonly referred to as ``Cosmic Noon" (redshift $z\sim2$), star-formation and gas accretion rates were significantly higher than in the local universe \citep[e.g.,][]{Madau2014, Tacconi2020}. The physical conditions of the interstellar medium (ISM) in galaxies at this epoch differ from those in the present-day universe, with notable variations in chemical composition, ionization states, and gas dynamics. Intense star formation in these systems can drive large-scale gas flows, one of the key feedback mechanisms that regulates galaxy evolution \citep[e.g.,][]{FS2018}. The rest-frame optical spectrum serves as a valuable probe of the ISM. It contains recombination lines such as \halpha{} and \hbeta{} and collisionally excited lines, including \oiiix{}, \niix{}, and [S\thinspace{\sc ii}]. These features trace physical properties of the ionized ISM including dust content, ionization conditions, and gas-phase metallicity.

A key diagnostic for understanding the ionization properties of galaxies is the Baldwin-Phillips-Terlevich (BPT) diagram \citep{Baldwin1981}, which empirically separates star-forming galaxies from those hosting active galactic nuclei (AGN) based on strong emission-line ratios. When applied to integrated galaxy spectra, the BPT diagram has been widely used to classify ionization sources. However, high-redshift star-forming galaxies exhibit a systematic offset from the local star-forming sequence, where \oiiihb{} and \niiha{} ratios are elevated \citep[e.g.,][]{Steidel2014, Strom2017, Runco2021, Shapley2025}. This systematic offset can push star-forming galaxies towards regions typically populated by AGN with low mass or high specific star formation rate host galaxies \citep[][]{Coil2015}. The physical origin of the offset in the star-forming sequence was previously under debate, with proposed explanations including higher nitrogen-to-oxygen (N/O) ratios at fixed oxygen abundance \citep[e.g.,][]{Masters2014}, harder ionizing spectra from massive stars \citep[e.g.,][]{Shapley2015, Sanders2023}, and higher ionization parameters \citep[e.g.,][]{Bian2020}. However, a preponderance of recent data now points toward harder ionizing spectra at fixed metallicity as the primary cause \citep[e.g.,][]{Steidel2016,Cullen2021,Shapley2025}

Significant progress has been made in characterizing high-redshift galaxies through decades of spectroscopic studies, providing key insights into their physical properties and evolution. Near-IR spectrographs on ground-based 8--10 m telescopes and the Hubble Space Telescope (HST) have enabled the spectroscopic study of galaxies up to $z\sim3$ \citep[e.g.,][]{Steidel2014,Shapley2015,Momcheva2016, Kashino2017,FS2019}. Recently, the redshift frontier has been extended significantly further with the advent of JWST, enabling rest-frame optical spectroscopy out to $z\sim11-14$ and beyond \citep[e.g.,][]{Carniani2024,Napolitano2025}. This extensive sample of galaxies across a wide range of redshifts provides statistical leverage on global properties such as metallicity, ionization conditions, and AGN activity. However, the majority of these studies rely on integrated spectra, where emission from different spatial regions is blended within a single aperture (but see, e.g., \citet{Jones2024}). This blending obscures the contribution of distinct ionization sources and spatially varying physical conditions, limiting the ability to fully disentangle the processes driving galaxy evolution.

Integral field unit (IFU) spectroscopy thus provides a crucial advantage by resolving spatial variations in emission-line ratios, allowing for a more nuanced interpretation of ionization conditions. By resolving spatial variations in line ratios and gas kinematics, IFU data can alleviate degeneracies and separate distinct ionization sources (e.g. star-forming regions, AGN, shocks) and kinematic components that may be mixed in integrated spectra \citep[e.g.,][]{FS2018,Ubler2024}. Spatially resolved BPT diagnostics enable the identification of weak or obscured AGN that may influence central emission-line ratios in predominantly star-forming galaxies \citep[e.g.,][]{Wright2010}, as well as the mapping of ionization structure within galaxies to constrain the underlying physical mechanisms \citep[e.g.,][]{Jones2024,Parlanti2025}.

Another key spatial diagnostic is the radial trend in gas-phase metallicity, often traced by strong-line ratios such as the \niiha{} flux ratio. In the local universe, star-forming disk galaxies exhibit a mild negative metallicity gradient ($\sim -0.05$~dex~kpc\sups{-1}), consistent with inside-out galaxy growth \citep[e.g.,][]{Rupke2010, Sanchez2014}. At Cosmic Noon, however, metallicity gradients are observed to be largely flat \citep[e.g.,][]{Curti2020, Wang2020, Ju2025, Li2025}, suggestive of efficient radial mixing due to strong feedback processes \citep[e.g.,][]{Gibson2013, Ma2017}. A subset of high-redshift galaxies show either negative \citep{Jones2010,Jones2013} or even inverted gradients, the latter potentially indicative of metal-poor gas accretion \citep[e.g.,][]{Wang2019, Li2022}. On the theoretical side, cosmological simulations with smooth stellar feedback produce strongly negative metallicity gradients at high-redshifts \citep[e.g.,][]{Garcia2025}, which is in contention with the flat gradients from observations. The large scatter in observed metallicity gradients and the mismatch with simulations point to a diverse range of physical mechanisms shaping galaxy evolution and ISM physical conditions during this epoch.

In this study, we present Keck/OSIRIS IFU observations of four galaxies at $z\sim2$, leveraging both high spectral ($R\sim3600$) and spatial resolution to investigate spatial variations in strong emission-line ratios. The largest sample of spatially resolved high-redshift galaxies to date comes from \citet{Wang2020}, based on low-resolution HST grism spectroscopy ($R\sim100$). Other ground-based IFU studies have also contributed to increase the sample size \citep[e.g.,][]{FS2019,Curti2020}, though observations are seeing-limited or are restricted to lensed galaxies. Our observations thus build on this foundation by utilizing the higher spectral resolution of OSIRIS to enable detailed multi-component emission-line fitting, as well as adaptive optics (AO) to achieve higher spatial resolution. Our work complements existing studies by providing new constraints on ionization conditions, AGN activity, and metallicity distributions within these galaxies, ultimately broadening our understanding of stellar feedback, outflows, and galaxy evolution.

The structure of this paper is as follows: in Section~\ref{sec:data}, we describe the observations and data reduction; in Section~\ref{sec:analysis}, we outline our methodology for spatial binning and emission-line fitting; in Section~\ref{sec:results}, we present results on BPT diagnostics and radial \niiha{} gradients; and in Section~\ref{sec:discussion}, we discuss the implications of our findings in the context of galaxy evolution.

Throughout this paper, we assume a $\Lambda$CDM cosmology with $H_0 = 70$ km s\sups{-1} Mpc\sups{-1}, $\Omega_m = 0.3$, and $\Omega_{\Lambda} = 0.7$.

%%%%%%%%%%%%%%%%%%%%%%%%%%%%%%%%%%%%%%%%%%%%%%%%%%%%%%%%%%%%%%%%%%%%%%%%%%%%%%%%%%%%%%%%%%%%%%%%%%
%%%%%%%%%%%%%%%%%%%%%%%%%%%%%%%%%%%%%%%%%%%%%%%%%%%%%%%%%%%%%%%%%%%%%%%%%%%%%%%%%%%%%%%%%%%%%%%%%%
\section{Observations \& data reduction} \label{sec:data}

\subsection{Targets \& observations} \label{subsec:targets_obs_redux}

The four galaxies analyzed in this study are drawn from the MOSFIRE Deep Evolution Field (MOSDEF) survey \citep{Kriek2015}, which provides rest-frame optical spectroscopy for a sample of $\sim 1500$ galaxies spanning a redshift range of $1.4 < z < 3.8$. These particular targets are among the brightest emission-line galaxies within the survey, and thus were selected for OSIRIS follow-up with the goal of detecting key diagnostic emission lines such as \hbeta, \oiii, \halpha, and \nii{} with high signal-to-noise. At the redshifts of these galaxies $(z \sim 2)$, these emission lines fall within the H and K bands.

Furthermore, three out of the four galaxies displayed signatures of gas flow from their MOSDEF spectra. As revealed by previous works, COSMOS~19985 and COSMOS~20062 have clearly-detected broad kinematic components in the strong emission lines listed above \citep[][]{Leung2017}, and GOODS-S~40768 shows a redshifted kinematic component indicative of gas infall \citep[][]{Weldon2023}.

A key consideration in target selection was the integrated flux of \halpha{}, which is typically the brightest rest-optical emission line within the MOSDEF samples. Sufficiently bright integrated \halpha{} flux $(\geq 1.5\times10^{-16}$~erg~s\sups{-1}~cm\sups{-2}$)$ is needed for reliable detection of fainter emission lines such as \niir{} and \hbeta{} given the typical line ratios observed in $z \sim 2$ star-forming galaxies. Additionally, each galaxy was required to have a tip-tilt star with an R-band magnitude of $R = 17$ or brighter within a $60''$ radius, enabling the use of adaptive optics (AO) correction to achieve $\sim0.1$'' resolution. Given these selection criteria, our subsample populates the more massive and higher SFR regime of the parent MOSDEF sample. The stellar masses and SFRs of our four galaxies are listed in Table~\ref{tab:obs_details}, where the median values of the $z\sim2$ MOSDEF sample are $10^{9.88}$~M$_{\odot}$ and 29~M$_{\odot}$~yr$^{-1}$ \citep[e.g.,][]{Shapley2022}.

Each of these galaxies also has corresponding rest-frame UV spectroscopic observations obtained with Keck/LRIS, providing complementary constraints on their ionizing radiation fields and extended gas flow kinematics \citep{Topping2020,Weldon2022}. With the ability of the OSIRIS IFU data to provide spatially resolved measurements of the gas kinematics, this sample presents a wealth of information on the excitation properties and physical processes in galaxies at $z\sim2$. A detailed kinematic analysis of these same OSIRIS observations is currently being carried out and will be presented in a separate, forthcoming paper.

Observations were conducted using the OH-Suppressing Infra-Red Imaging Spectrograph (OSIRIS) \citep{OSIRIS2006} on the Keck I telescope, which is an integral field unit (IFU) operating in the near-infrared (NIR). OSIRIS, when used with the Keck adaptive optics system, enables spatially resolved spectroscopy at resolutions of $\sim 0.1''$, which corresponds to $\sim 1$~kpc at $z \sim 2$. This high angular resolution is critical for studying the spatial variations in emission-line properties across each galaxy.

The four OSIRIS target galaxies were observed on five different nights spanning May 2017, January 2018, and June 2018. Each galaxy was observed in both the H narrow band (Hn2 or Hn3) and K broad band (Kbb) using a lenslet plate scale of $0.05''$ and with Laser Guide Star (LGS) AO. Each individual frame was exposed for 900 seconds to give a total exposure time that ranged from 5400--12600 seconds in the H bands and 5400--10800 seconds in the K bands. A dither of 1.6'' in the long axis of the FOV was applied between exposures to enable sky subtraction. Information about the observations is listed in Table~\ref{tab:obs_details}.

\begin{deluxetable*}{l|cccccccc}
\tablecaption{Observation Details \label{tab:obs_details}}

\tablehead{
    \colhead{ID} &
    \multicolumn{2}{c}{COSMOS~19985} &
    \multicolumn{2}{c}{COSMOS~20062} &
    \multicolumn{2}{c}{AEGIS~3668} &
    \multicolumn{2}{c}{GOODS-S~40768}
}

\startdata
% ID                                                  & \multicolumn{2}{c}{COSMOS~19985}      & \multicolumn{2}{c}{COSMOS~20062}      & \multicolumn{2}{c}{AEGIS~3668}      & \multicolumn{2}{c}{GOODS-S~40768}      \\
R.A (J2000)                                         & \multicolumn{2}{c}{10:00:14.484}      & \multicolumn{2}{c}{10:00:16.436}      & \multicolumn{2}{c}{14:19:28.839}    & \multicolumn{2}{c}{03:32:09.797}       \\
Dec. (J2000)                                        & \multicolumn{2}{c}{+02:22:57.98}      & \multicolumn{2}{c}{+02:23:00.79}      & \multicolumn{2}{c}{+52:46:24.62}    & \multicolumn{2}{c}{-27:43:08.65}       \\
$z$                                                 & \multicolumn{2}{c}{2.1882}            & \multicolumn{2}{c}{2.1857}            & \multicolumn{2}{c}{2.1877}          & \multicolumn{2}{c}{2.3035}             \\
$\log(\mathrm{M}_{\ast}/\mathrm{M}_{\odot})$        & \multicolumn{2}{c}{$10.10_{-0.13}^{+0.09}$}             & \multicolumn{2}{c}{$10.26_{-0.07}^{+0.03}$}             & \multicolumn{2}{c}{$9.93_{-0.06}^{+0.11}$}            & \multicolumn{2}{c}{$9.79_{-0.08}^{+0.02}$}               \\
SFR $(\mathrm{M}_{\odot}\,\mathrm{yr}^{-1})$ & \multicolumn{2}{c}{$193_{-11}^{+12}$}                 & \multicolumn{2}{c}{$87_{-44}^{+79}$}                 & \multicolumn{2}{c}{$41_{-6}^{+7}$}               & \multicolumn{2}{c}{$216_{-15}^{+16}$}                  \\
$H_{160}$ (AB mag)                                  & \multicolumn{2}{c}{21.87}             & \multicolumn{2}{c}{22.10}             & \multicolumn{2}{c}{22.35}           & \multicolumn{2}{c}{22.04}              \\
\cline{2-9}
Filter                                              & Hn2               & Kbb               & Hn2               & Kbb               & Hn2         & Kbb                   & Hn3                & Kbb               \\
\multirow{2}{*}{Date (MM/DD/YY)}                                     & \multirow{2}{*}{01/02/18}          & \multirow{2}{*}{01/02/18}          & \multirow{2}{*}{01/03/18}          & \multirow{2}{*}{01/03/18}          & \multirow{2}{*}{05/17/17}    & 05/18/17,    & \multirow{2}{*}{01/02/18}           & \multirow{2}{*}{01/03/18}          \\
                                     &           &           &           &           &     & 06/05/18    &            &           \\
$t_\mathrm{exp}$ (s)                                & 5400              & 9000              & 7200              & 5400              & 7200        & 10800                 & 12600              & 5400              \\
PSF FWHM ($''$)                                          & 0.130             & 0.123             & 0.140             & 0.142             & 0.114       & 0.117                 & 0.109              & 0.097             
\enddata
\end{deluxetable*}

%----------------------------------------------------%
\subsection{Data reduction} \label{subsec:reduction}

The observations were reduced using the OSIRIS Data Reduction Pipeline \citep{DRPLyke2017,DRPLockhart2019}, with several modifications to the standard reduction procedure due to the absence of a dedicated sky frame.

Initially, a bad pixel mask was applied to correct for detector defects, followed by the subtraction of a master dark that was generated from median-combining several dark frames observed at the start of the night. The pipeline then applied previously-measured lenslet spectral-PSF maps to construct the datacube using a process similar to Lucy-Richardson deconvolution. 

Sky subtraction was achieved by using the pipeline's scaled sky subtraction procedure, which measures the flux of the brightest OH emission lines to determine the scaling of the corresponding families of lines for sky-line subtraction. In lieu of a dedicated sky frame, we applied the procedure on pairs of dithered images by subtracting each from the other. The resulting frames were then mosaicked using a sigma clipping algorithm.

To estimate the Point Spread Function (PSF) of the data, the same process was applied to the observations of the tip-tilt stars associated with each galaxy. We fitted a symmetric 2D-Gaussian to the flux map obtained by summing up the flux within each spatial pixel (spaxel) and defined the PSF of the observations as the Full Width at Half Maximum (FWHM) of the 2D-Gaussian model. These values are quoted in Table~\ref{tab:obs_details}, with an average of $0.123''$ in the H bands and $0.120''$ in the K band.

During the reduction process, it became evident that the error spectra generated by the pipeline were elevated relative to the noise estimated empirically in parts of the spectra between strong OH emission lines. To address this inconsistency and scale down the pipeline error spectra, we first defined wavelength windows in regions that lacked significant OH line emission, with observed widths of $\sim 30$~\AA{} in the H band and $\sim 50$~\AA{} in the K band. We then calculated the ratio between the 1$\sigma$ standard deviation of the flux and the median of the pipeline error spectrum within these windows for each spaxel, taking the median ratio across the wavelength windows in each spaxel to obtain a map of error-scaling factors. The median value of this error-scaling factor map was then applied uniformly to rescale the pipeline error spectrum across all spaxels.

Additionally, the OH emission lines in the raw science frames appeared to be offset from their expected wavelengths based on the known wavelengths of near-IR OH sky lines. We corrected for this offset by performing centroiding on the OH emission lines in a dark-subtracted, non-sky-subtracted science data cube, calculating the mean offset per spaxel, and applying mean value of all these offsets to shift the wavelength calibration of the final reduced data. The averages of these offsets are $-2.18 \pm 0.04$~\AA{} and $-3.29 \pm 0.05$~\AA{} in the H- and K-bands for the 2018 observations (i.e. COSMOS~19985, COSMOS~20062, and GOODS-S~40768), and $\sim0.32$~\AA{} and $\sim0.53$~\AA{} in the H- and K-bands for the 2017 observations (i.e. AEGIS~3668).

As the analyses presented in this work only involve relative emission-line flux ratios, we do not perform flux calibration in physical units to the data. For readers interested in the calibrated emission-line fluxes and derived fundamental properties for these systems, we refer to the MOSDEF survey results presented in \citet{Kriek2015}, \citet{Shapley2022}, and subsequent MOSDEF papers.

%%%%%%%%%%%%%%%%%%%%%%%%%%%%%%%%%%%%%%%%%%%%%%%%%%%%%%%%%%%%%%%%%%%%%%%%%%%%%%%%%%%%%%%%%%%%%%%%%%
%%%%%%%%%%%%%%%%%%%%%%%%%%%%%%%%%%%%%%%%%%%%%%%%%%%%%%%%%%%%%%%%%%%%%%%%%%%%%%%%%%%%%%%%%%%%%%%%%%
\section{Analysis} \label{sec:analysis}

\subsection{Spatial binning} \label{subsec:binning}

To obtain robust emission-line maps that are based on line ratios spanning the H and K bands, it is essential to align the data cubes to ensure that we are comparing the same physical regions of the galaxy across both bands. To achieve this alignment, we first created flux maps for a 10~\AA{} window centered around the strongest emission lines in each band (\oiiir{} for H band and \halpha{} for K band). These maps were normalized to the brightest pixel near the galaxy’s center. To align the bands, we computed the squared residuals between the flux maps, shifting the H band flux map relative to the K band flux map iteratively on integer spaxel increments and identifying the configuration that minimized the residuals. Lastly, we cropped the data in both bands to the same size, which is typically limited by the FOV of the K-band data.

We applied two types of spatial binning methods for our subsequent analysis. First, we performed a coarser radial or circular aperture binning on both H and K bands for the BPT diagram analysis. Second, we applied a finer Voronoi binning based on the higher S/N data in the K band to enable a more detailed examination of the \niiha{} ratio. These binning strategies are described in detail in the following sections.

%----------------------------------------------------%
\subsubsection{Radial and clump binning} \label{subsubsec:radial_binning}

For galaxies COSMOS~19985, COSMOS~20062 and AEGIS 3668, radial binning was employed to investigate spatial variations in emission line properties with increasing distance from the center of the objects. A 2D Gaussian was fit to the higher-S/N K-band flux map of each galaxy to parameterize the elliptical morphology, with key parameters such as semi-major and semi-minor axes containing the central 68.3\% of the flux (i.e., one spatial standard deviation), orientation, and centroid. Using this parameterization, we extracted regions encompassing up to 99.7\% of the flux (i.e., 3 spatial standard deviations) and generated a mask to isolate the galaxy in both bands. Radial bins were then constructed, with a central bin encompassing 1 spatial-$\sigma$ of the flux and additional annuli encompassing 2 spatial-$\sigma$ and 3 spatial-$\sigma$ flux levels. These bins are named \Rsig{1}, \Rsig{2}, and \Rsig{3} respectively. The radial bin map for each galaxy is given in the bottom-left panel of Figs.~\ref{fig:flux_bin_maps_co19985}-\ref{fig:flux_bin_maps_aeg905}.

Due to the unusual morphology of GOODS-S~40768 consisting of three distinct clumps, instead of radial binning, we defined three circular apertures centered on regions exhibiting similar kinematics. The center, East, and West bins shown in Figs.~\ref{fig:flux_bin_maps_gs40768} have radii 0.175$''$, 0.15$''$, and 0.125$''$ respectively. Another clump with higher \halpha{} surface brightness is identified to the upper-left of the center of COSMOS~20062, so an additional circular aperture of radius 0.1$''$ was defined to isolate this distinct clump.

For each bin, the mean spectrum was calculated by averaging the flux across all spaxels within the bin. The error spectrum was summed in quadrature across the spaxels and divided by the square root of the number of spaxels in the bin, yielding a representative uncertainty for the mean spectrum.

%----------------------------------------------------%
\subsubsection{Voronoi binning} \label{subsubsec:vorbin}

In the case of joint analyses of H- and K-band data, the spatial resolution of the radial bins described above is limited by the lower S/N in the H band. Since the examination of radial variations in \niiha{} ratios only depends on K-band data, we adopt finer bins using Voronoi binning for this higher S/N data to enhance the spatial resolution of this analysis. Voronoi binning was applied using the Python package \texttt{vorbin} \citep{Cappellari2003}. This method takes a S/N map and groups spaxels together to reach a target S/N threshold while striving to make the bins as round as possible, thereby preserving maximum spatial resolution. For this analysis, the binning was performed using the \halpha{} flux map and the 3 spatial-$\sigma$ elliptical mask described above. The associated error map was constructed as the 1$\sigma$ standard deviation in flux values from neighboring continuum wavelength regions. Spaxels with S/N $< 1$ were excluded from further analysis.

Voronoi binning was applied to COSMOS~19985, COSMOS~20062, and AEGIS~3668. The object GOODS-S~40768 was excluded from this method since the clump bins in Section~\ref{subsubsec:radial_binning} already describe the spatial variations well and low S/N prevents additional bins from being created. The target S/N threshold for the two COSMOS objects were set such that the \niir{} emission line reaches a minimum of $S/N \sim 3$ in every bin. Due to the low \niir{} flux in AEGIS~3668, we imposed a less stringent limit on its \niir{} S/N. Instead, the target S/N of \halpha{} for Voronoi binning of AEGIS~3668 was set to be comparable to that of the two COSMOS objects, which had target $S/N \sim 25$. 

\begin{figure*}[ht!]
    \centering
    \includegraphics[width=\textwidth]{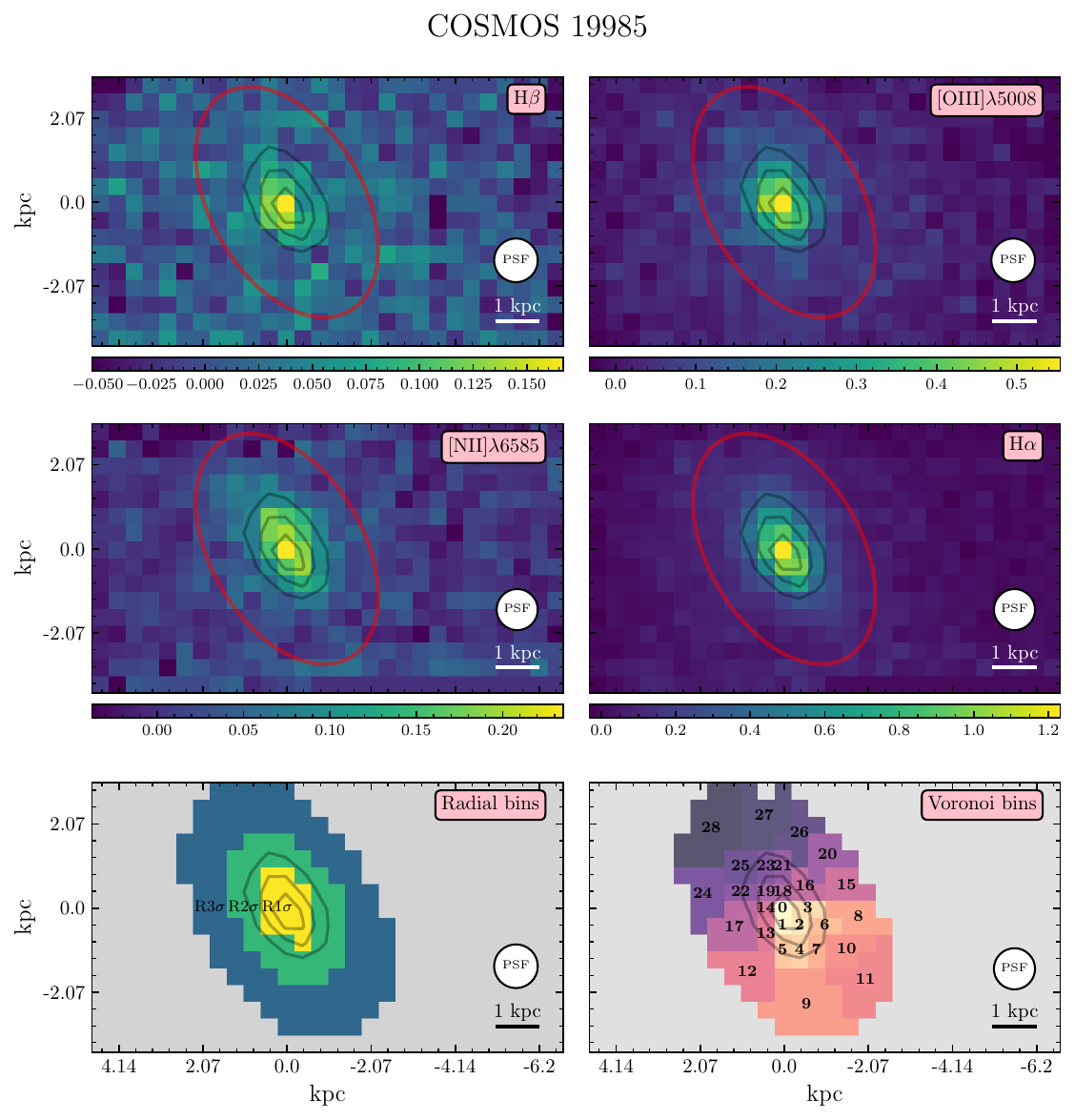}
    \caption{Emission line flux maps and spatial binning maps for COSMOS~19985. The upper four maps represent the summed flux in a rest-frame 10~\AA{} window centered around \hbeta, \oiiir, \niir, and \halpha{} with arbitrary flux units. Gray contours that trace the fixed \halpha{} flux levels are overlaid on top of each map for reference. The red ellipses mark the 3-$\sigma$ flux-level threshold of the galaxy based on a 2D-Gaussian model. The bottom left map shows the radial bins generated from the 1-$\sigma$, 2-$\sigma$, and 3-$\sigma$ flux-level thresholds, while the bottom right map shows the Voronoi bins. Bin colors are added for clarity. Circles indicating the FWHM PSF are overlaid on each map. The PSF of the radial bins is given by the larger PSF between the H and K band. The PSF of the Voronoi bins is set by that of the K band.} 
    \label{fig:flux_bin_maps_co19985}
\end{figure*}

\begin{figure*}[ht!]
    \centering
    \includegraphics[width=\textwidth]{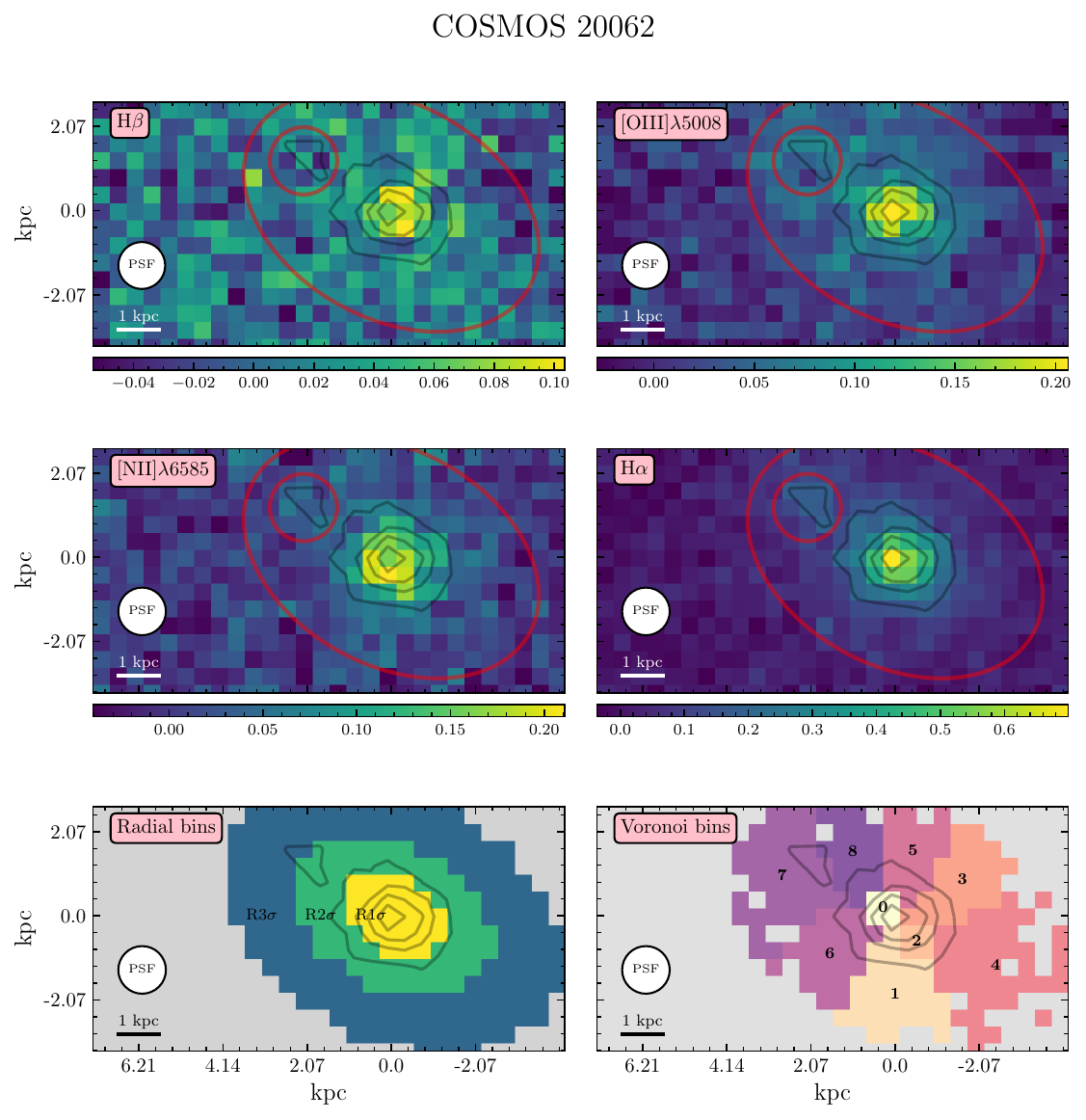}
    \caption{Emission line flux maps and spatial binning maps for COSMOS~20062, as in Fig.~\ref{fig:flux_bin_maps_co19985}. The additional small red circle denotes the clump bin with radius.
    \label{fig:flux_bin_maps_co20062}}
\end{figure*}

\begin{figure*}[ht!]
    \centering
    \includegraphics[width=\textwidth]{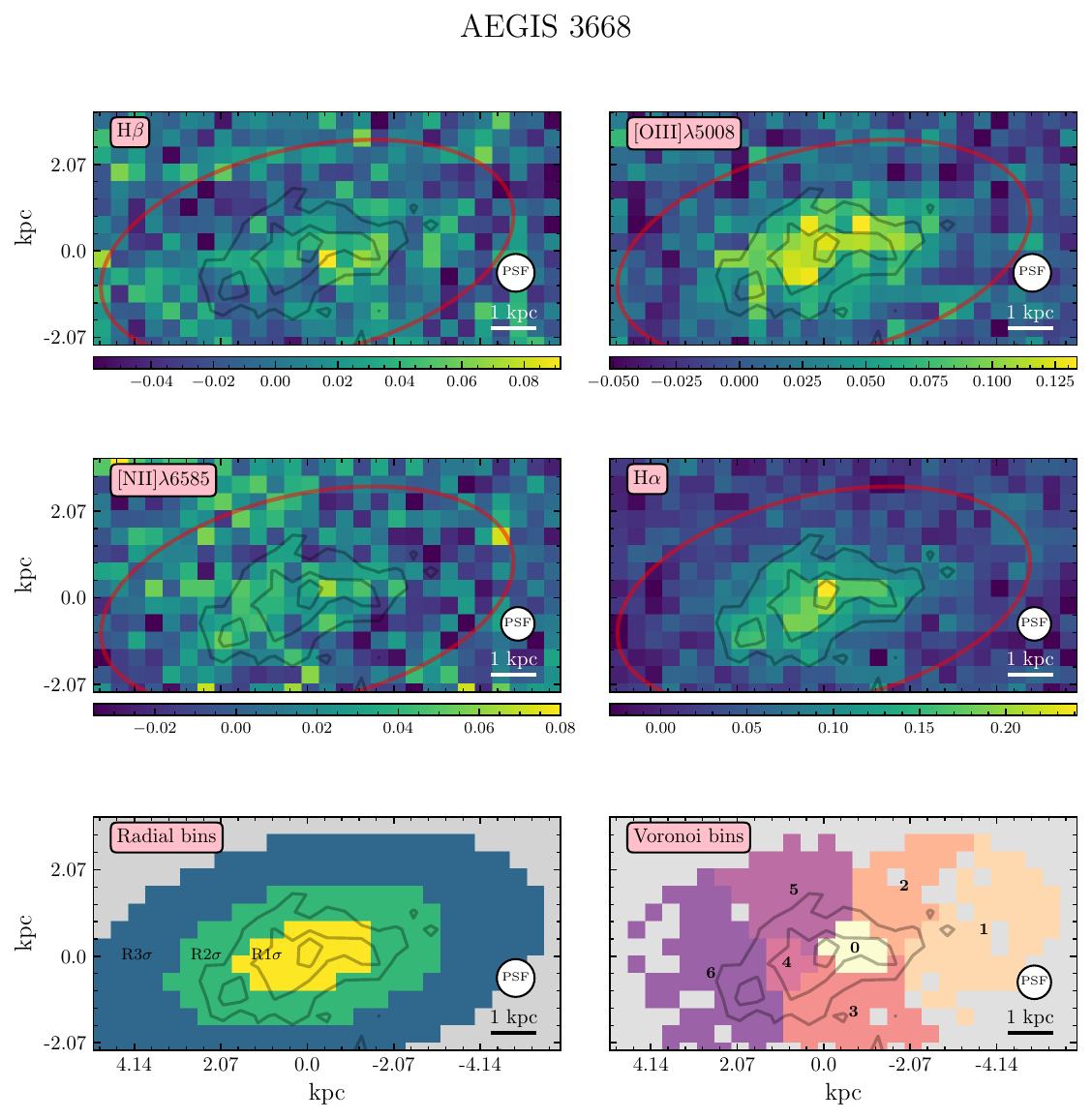} 
    \caption{Emission line flux maps and spatial binning maps for AEGIS~3668, as in Fig.~\ref{fig:flux_bin_maps_co19985}.
    \label{fig:flux_bin_maps_aeg905}}
\end{figure*}

\begin{figure*}[ht!]
    \centering
    \includegraphics[width=\textwidth]{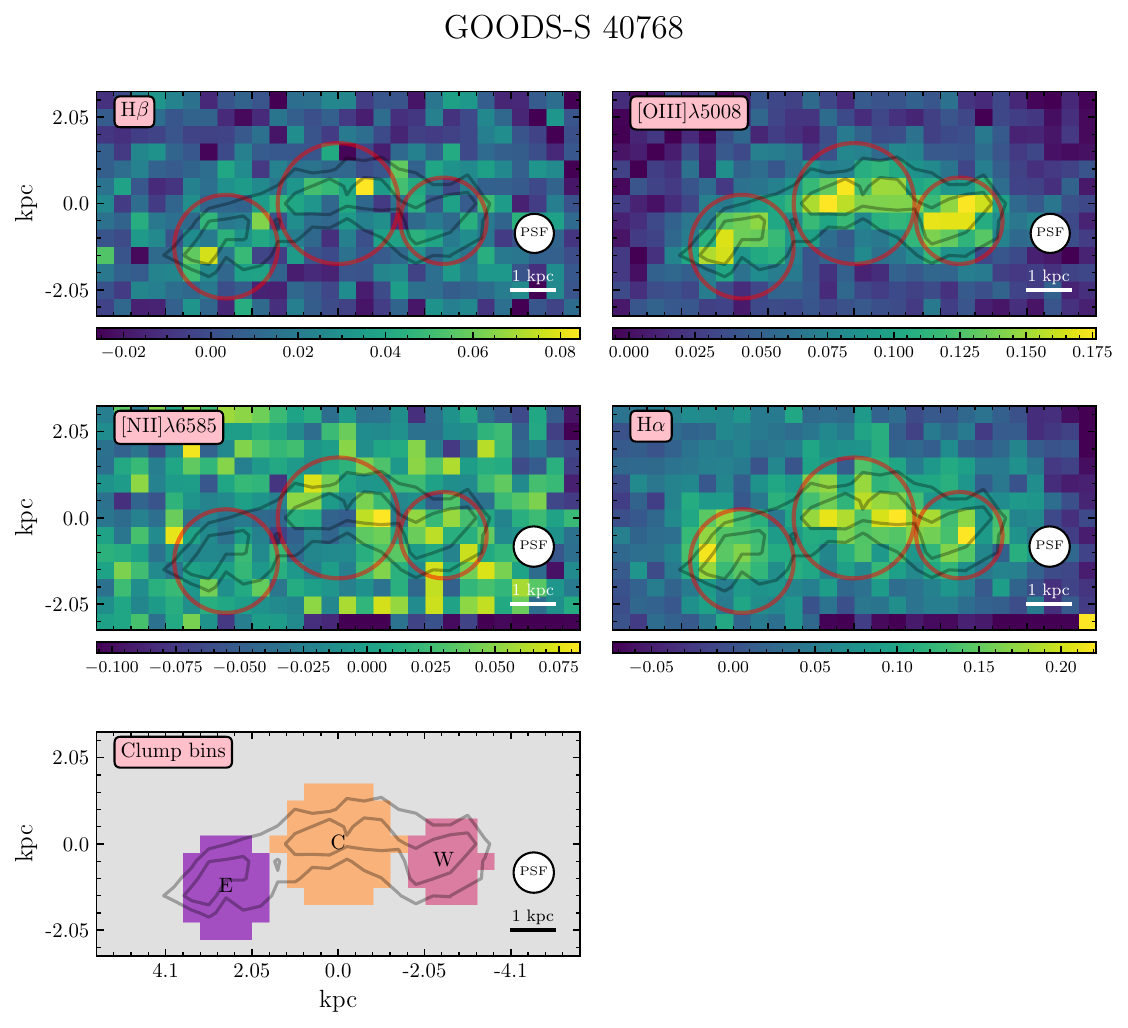} 
    \caption{Emission line flux maps and spatial binning maps for GOODS-S~40768, as in Fig.~\ref{fig:flux_bin_maps_co19985}. Here, the gray contours trace the higher S/N \oiiir{} fixed flux levels. Only three clump bins are defined for this galaxy due to its unusual morphology and lower S/N.
    \label{fig:flux_bin_maps_gs40768}}
\end{figure*}

Figures~\ref{fig:flux_bin_maps_co19985}--~\ref{fig:flux_bin_maps_gs40768} show the maps of the spatial binning strategies as described above. We note that some of the Voronoi bins, for example the central bins in COSMOS~19985, are smaller than the PSF of the observation. As such, measurements made in each of these bins are likely correlated to those of adjacent bins. Therefore, we caution against overinterpreting individual measurements in these bins. This limitation is further discussed in Section~\ref{subsec:radial_niiha}.

%----------------------------------------------------%
\subsection{Spectral analysis} \label{subsec:spec_analysis}

\subsubsection{Emission line fitting} \label{subsubsec:emline_fitting}

Emission line fitting was performed using the Python package \texttt{lmfit} to model the spectral features of each galaxy. We fit the \nii+\halpha{} and \oiii+\hbeta{} emission lines simultaneously within their respective groups using Gaussian profiles parameterized by flux, Doppler shift, and broadening, each corresponding to the amplitude, center, and width of a Gaussian.

For each group, we tied the Doppler shift and broadening width of the weaker lines to those of the strongest line in the group -- \halpha{} for the \nii+\halpha{} group and \oiiir{} for the \oiii+\hbeta{} group. The flux ratios of the \nii{} doublet and the \oiii{} doublet were fixed to their theoretical values of 2.94 and 2.98, respectively.

After enforcing the above relationships, we ended up with four key fitting parameters for each group:
\begin{enumerate}
    \item Center of the strongest line (\oiiir{} or \halpha)
    \item Broadening width of the strongest line
    \item Flux of the hydrogen recombination line (\halpha{} or \hbeta)
    \item Flux of the stronger metal line (\niir{} or \oiiir)
\end{enumerate}

We applied two types of fits to each group of emission lines. The one-component model used a single Gaussian per emission line, resulting in a total of three Gaussians per group and four fitting parameters. In contrast, the two-component model had two Gaussians per emission line: a narrow component (C1) representing systemic gas and a broad component (C2) representing gas flows. The line centers for C1 and C2 were allowed to differ. This approach resulted in a total of six Gaussians per group and eight fitting parameters. In the two-component model, the narrow component (C1) was constrained to have a narrower width, tracing the local velocity dispersion in the ionized ISM, while the broad component (C2) captured possible signatures of gas flow. We treat the fitted Gaussians in the one-component model as a narrow component.

Additional models were also employed in response to poor fits to certain spectra from visual inspection. Details about these more complex models and the spectrum that was best fit by one of such models will be discussed in Section~\ref{subsubsec:spec_co19985}.

To ensure physically reasonable fits, we imposed limits on the velocity shifts and broadening parameters. The bounds were chosen to be loose enough to prevent the fit from adopting the bound values but strict enough to be physically reasonable. For one-component fits, we allowed a maximum velocity shift of $\pm 500$ km~s\sups{-1} and a maximum FWHM of 2000 km~s\sups{-1}. In two-component fits, we introduced a more restrictive narrow component with velocity shift limit of $\pm 100$ km~s\sups{-1} and maximum FWHM of 1000 km~s\sups{-1}. Additionally, the minimum FWHM was set based on the spectral resolution of $R \sim 3600$, and flux values were restricted to be non-negative.

Fitting errors were estimated using the square root of the diagonal terms of the covariance matrix. To avoid overfitting with the two-component model, we used the Bayesian Information Criterion \citep[BIC;][]{KassRaftery1995} to assess model preference. For the more complex two-component model to be favored, we required its BIC value to be at least 6 lower than that of the one-component model, indicative of strong evidence against the simpler model \citep{KassRaftery1995}; otherwise, the one-component model was selected.

\begin{figure*}[ht!]
    \centering
    \includegraphics[width=0.85\textwidth]{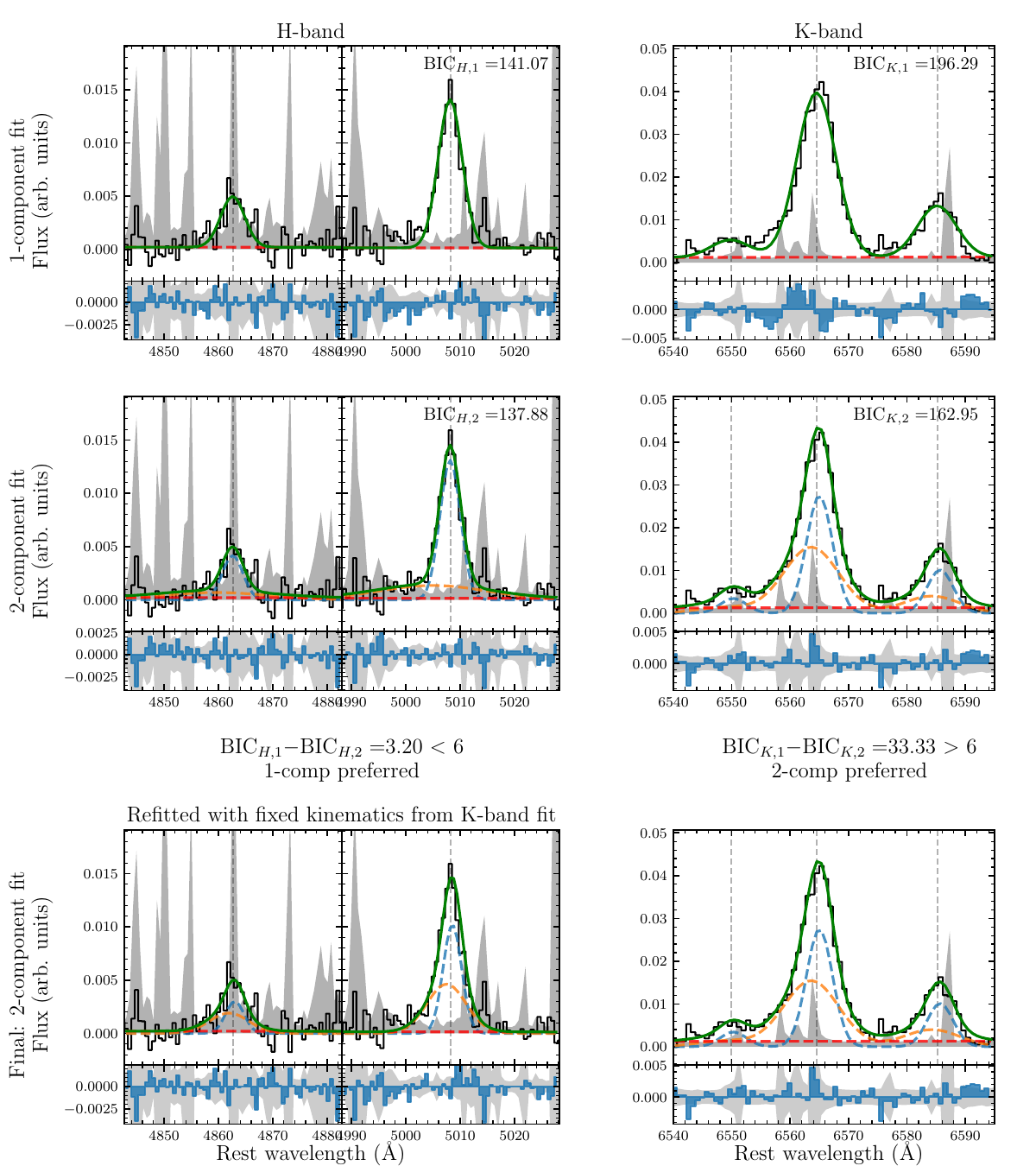} 
    \caption{Example of emission line fitting routine for radial bin \Rsig{1} of COSMOS~20062. The left column shows the H band spectrum with emission lines \hbeta{} and \oiiir{}, while the right column shows the K band spectrum with the \nii+\halpha{} emission lines. The solid black curves represent the flux spectra and the gray shaded regions represent the error spectra. Each spectrum is fit with both a one-component (1-comp) and two-component (2-comp) model. For radial and clump bins, this fitting is performed for both H and K bands, while for Voronoi bins, it is applied only to the K band. The green solid curve represents the total fit, while the red dashed line shows the linear continuum fit around the emission lines. In the 2-comp model, the blue dashed line corresponds to the narrow component, and the orange line represents the broad component. The Bayesian Information Criterion (BIC) is used to assess model preference, with the more complex 2-comp model favored if its BIC value is at least 6 lower than that of the 1-comp model; otherwise, the 1-comp model is selected. For radial and clump bins, the preferred models are compared between the H and K bands. If the preference differs, the model from the band with the higher S/N is adopted. The spectrum in the lower S/N band is then refitted with its kinematics (Doppler shift and line broadening) fixed to the fitted values of the higher S/N band. 
    \label{fig:co20062_example_fitting}}
\end{figure*}

For the integrated OSIRIS and radially-binned spectra, we further cross-checked the fitting between the H and K bands. In cases where the \nii+\halpha{} and \oiii+\hbeta{} groups disagreed on the preferred component fit (i.e.,  one- vs. two-component), refitting was performed based on the preferred component fit of the higher S/N data (typically the K band with \nii+\halpha). The S/N was calculated as the ratio of the flux to the fitting error of the strongest line in the group. During refitting, the velocity shift and broadening width were fixed to the fitted values from the higher S/N data, while only the fluxes were allowed to vary in the lower S/N data (typically the H band with \oiii+\hbeta). Fig.~\ref{fig:co20062_example_fitting} shows a schematic of the above fitting routine.

Since both groups of emission lines should trace roughly the same gas, we flagged fits where the velocity shift or broadening width in the lower S/N data (H band) differed from that in the higher S/N data (K band) by more than $1\sigma$ of the fit error to ensure consistency of the kinematics between the groups of emission lines. The lower S/N data was then refitted using the same kinematic-fixing procedure as above.

As a final quality check, we ensured that all fitted fluxes exceeded three times the fitting error ($3\sigma$). Spectra that did not meet this criterion were flagged and $3\sigma$ upper-limits were quoted in subsequent analysis.

For the integrated spectrum of all spaxels in some of the objects, the center of the narrow component did not match the systemic redshift determined from the spectra obtained from the MOSDEF survey. We performed a two-component fit to the integrated spectra and redefined a systemic redshift based on the fitted center of the narrow component before performing fitting on the binned spectra. Furthermore, to ensure consistency in comparisons between the MOSFIRE and integrated OSIRIS spectra, we required that the MOSFIRE spectra adopt the preferred component fit of the corresponding integrated OSIRIS spectra. The fits to the MOSFIRE spectra along with discussion on their preferred component fits are found in Appendix~\ref{app_subsec:mos_int}.

Examples of all the above spectra and fitted profiles can be found in the Appendix.

%----------------------------------------------------%
\subsubsection{COSMOS~20062 clump} \label{subsubsec:spec_co20062}

\begin{figure}[ht!]
    \centering
    \includegraphics[width=0.45\textwidth]{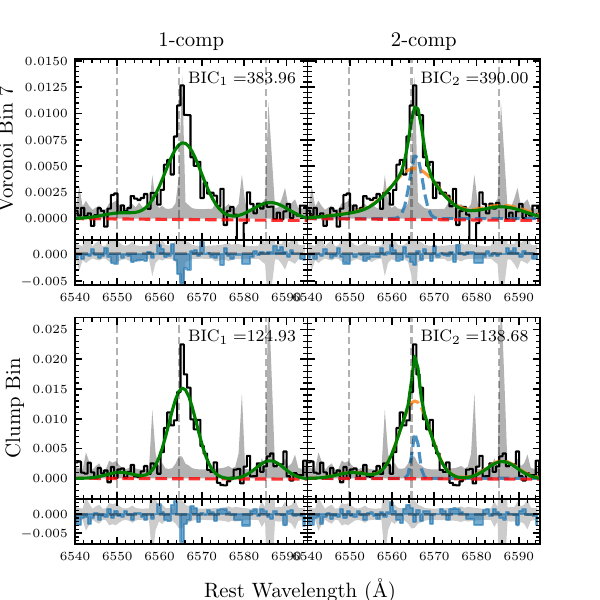} 
    \caption{Spectral fits for bins tracing the bright clump in COSMOS~20062. The top row shows Voronoi Bin 7, which connects the galaxy center to the clump, while the bottom row shows a circular aperture with 0.1$''$ radius centered on the clump. The left and right columns show the one- and two-component fits respectively. The color scheme follows that of Fig.~\ref{fig:co20062_example_fitting}. The BIC value for each fit is quoted at the upper-right corner of each panel. While the two-component model appears to fit better, the BIC analysis favors the one-component model, likely due to low S/N.
    \label{fig:co20062_clump_1v2comp_bic_comparison}}
\end{figure}

From the fits for bins tracing the bright clump in COSMOS~20062 (Fig.~\ref{fig:co20062_clump_1v2comp_bic_comparison}), we find large systematic residuals, particularly in Voronoi Bin 7, which extends from the galaxy center to the clump to capture a larger spatial region. Although the two-component model appears to better encapsulate the \nii+\halpha{} profiles, the lower BIC values for the one-component model in both cases indicate that the simpler model is preferred. This preference is likely due to the lower S/N and noise spikes near the peaks of \halpha{} and \niir{}, which limit the addition of more fitting parameters. 

Additionally, we employed the $\chi^2$ model selection method described in \citet{Leung2017}, in which the justification of a more complex model is determined using survival statistics that account for the differing numbers of degrees of freedom between the one- and two-component models. According to this criterion, neither Voronoi Bin~7 nor the clump bin yields a $p$-value $<0.01$ from the reduction in $\chi^2$, and both therefore prefer a one-component model.

While a more complex fit may be warranted by visual inspection, the available data do not provide strong statistical justification for its adoption. Subsequent analysis will proceed using the one-component fits.

%----------------------------------------------------%
\subsubsection{COSMOS~19985 central spaxel} \label{subsubsec:spec_co19985}

\begin{figure*}[ht!]
    \centering
    \includegraphics[width=\textwidth]{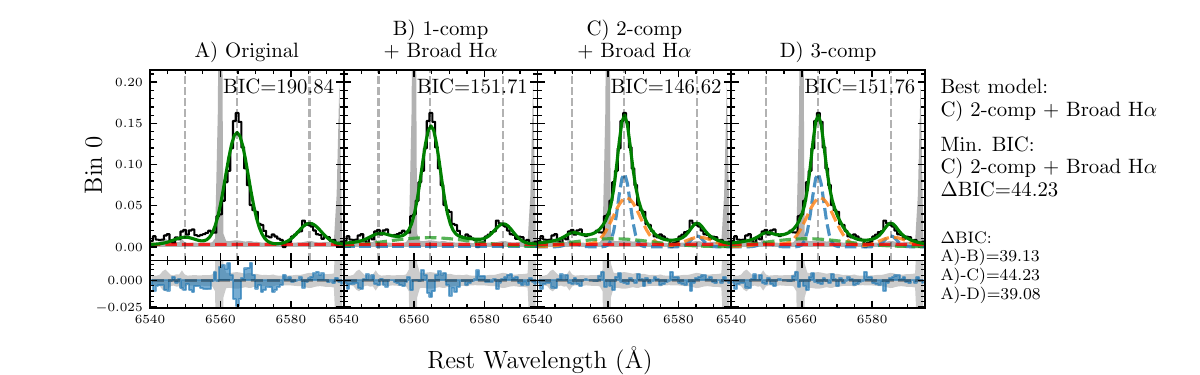} 
    \caption{Spectral fits for Voronoi Bin 0 (i.e. center-most spaxel) in COSMOS~19985. The color scheme follows that of Fig.~\ref{fig:co20062_example_fitting}. The BIC value of each fit is quoted at the upper-right corner of each panel. Model A) is the one-component model preferred by the routine described in Section~\ref{subsubsec:emline_fitting}. Models B) and C) are one- and two-component profiles with an additional broad \halpha{} Gaussian. Model D) is a three-component model. The best model, as determined by BIC analysis, is shown in the right column, along with the differences in BIC values between models.
    \label{fig:co19985_centerspax_bic_comparison}}
\end{figure*}

From Fig.~\ref{fig:co19985_centerspax_bic_comparison}, it is evident that the best-fit one-component model determined by the fitting routine does not provide an adequate fit to the emission-line profiles. Thus, we introduced additional models, including a one-component fit with a broad \halpha{} component, a two-component fit with a broad \halpha{} component, and a three-component fit. The best-fit model was chosen to be the model with the lowest BIC value, with a minimum reduction of $\Delta\mathrm{BIC}>6$ as compared to the original model to favor the new, more complex model. This analysis was conducted for all galaxies with Voronoi binning. Among them, only COSMOS~19985 exhibited spectra where one of these new models yielded the minimum BIC (see Fig.~\ref{fig:co19985_interesting_bic_comparison}). Even in this case, the preference for a more complex model was limited to Voronoi Bin 0 (i.e. the central-most spaxel), where the two-component fit with a broad \halpha{} component model was favored, as highlighted in Fig.~\ref{fig:co19985_centerspax_bic_comparison}. We quote measurements from the two-component plus broad \halpha{} fit in subsequent analysis.

%%%%%%%%%%%%%%%%%%%%%%%%%%%%%%%%%%%%%%%%%%%%%%%%%%%%%%%%%%%%%%%%%%%%%%%%%%%%%%%%%%%%%%%%%%%%%%%%%%
%%%%%%%%%%%%%%%%%%%%%%%%%%%%%%%%%%%%%%%%%%%%%%%%%%%%%%%%%%%%%%%%%%%%%%%%%%%%%%%%%%%%%%%%%%%%%%%%%%
\section{Results} \label{sec:results}

The integral field spectroscopic data obtained with OSIRIS provide crucial new insight into the spatial variations of galaxy properties. After fitting the emission-line profiles and applying two different binning methods, we can investigate how emission line flux ratios vary across different regions of each galaxy, allowing us to probe the nature of the ionization mechanisms shaping the ISM. In particular, we focus on the key emission lines \oiiir, \hbeta, \niir, and \halpha, examining how their flux is distributed between narrow and broad components and how these flux ratios change with radial distance. The following sections explore these spatial trends, leveraging emission-line diagnostics to gain insight into the ISM physical conditions within our galaxy sample.

%----------------------------------------------------%
\subsection{BPT diagrams} \label{subsec:BPT}

One of the most widely used emission-line diagnostic tool for identifying the ionization mechanism of the ISM is the relationship between \oiiihb{} and \niiha{}, introduced by \cite{Baldwin1981} (hereafter BPT). The placement of a galaxy on the BPT diagram helps determine whether its gas is photoionized by starlight, shocks, or an AGN.

First, we obtain the flux ratios \oiiihb{} and \niir/\halpha{} from our fitted emission line profiles in the radial and clump bins that contain all four required emission lines. Since each ratio consists of emission lines within the same spectral region, we mitigate the need for flux calibration between bands or dust correction. Additionally, by separating the narrow and broad components, we can examine how emission is distributed between these components and assess radial variations in the narrow component, which primarily traces systemic gas and the ISM.

\begin{figure*}[ht!]
    \centering
    \includegraphics[width=\textwidth]{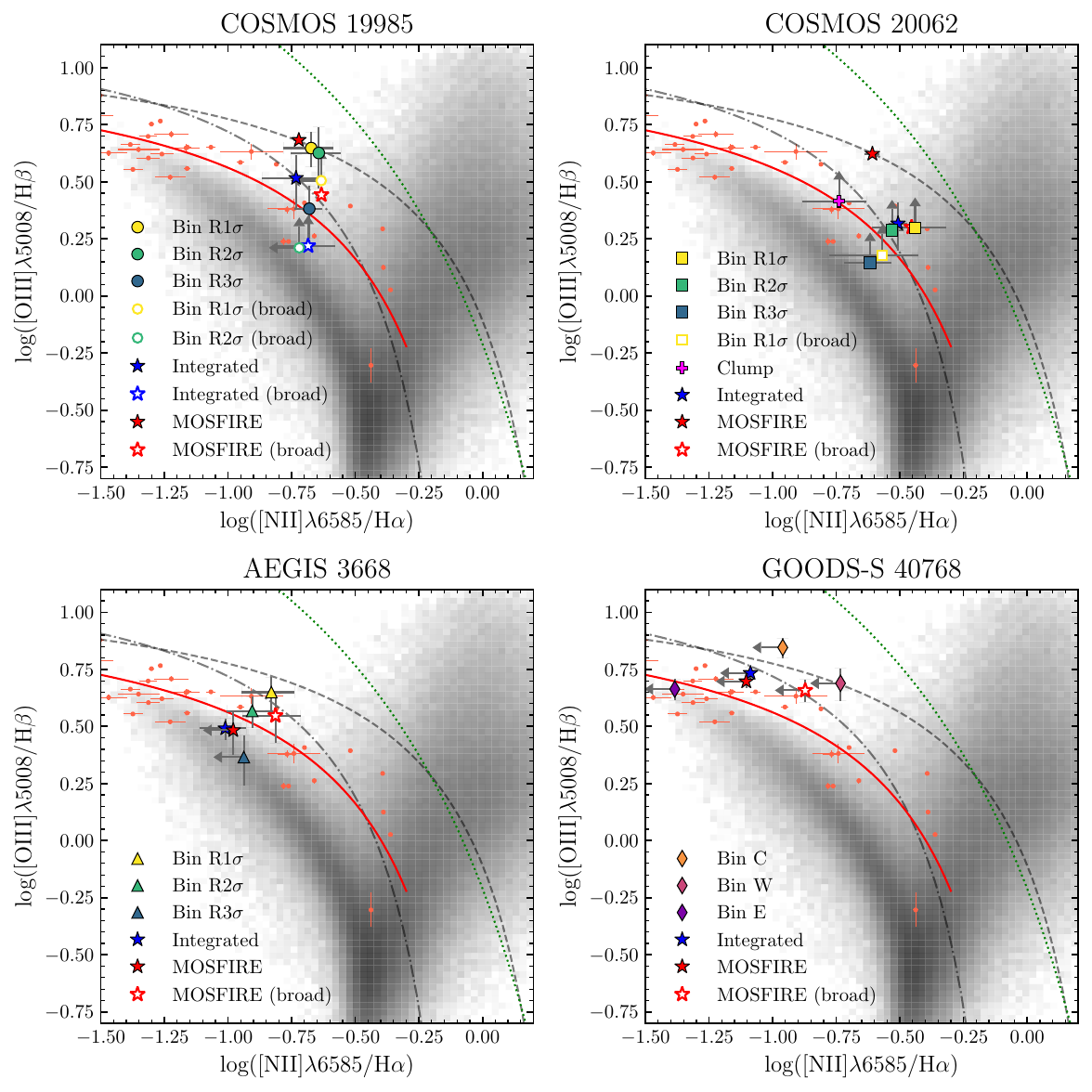} 
    \caption{BPT emission-line diagnostic diagrams for each galaxy: COSMOS~19985 (top left), COSMOS~20062 (top right), AEGIS~3668 (bottom left), and GOODS-S~40768 (bottom right). In each panel, the background grayscale histogram shows the distribution of local SDSS galaxies, the small red dots are $z=1.4-2.7$ galaxies from \cite{Shapley2025}, and the red curve denotes the median trend of $z\sim2$ MOSDEF galaxies \citep{Runco2021}. The dashed curve is the ``maximum starburst" line from \citet{Kewley2001}, while the dot-dashed curve is the empirical AGN/star-formation threshold at $z\sim0$ from \citet{Kauffmann2003} and the green dotted curve is the threshold for high-redshift galaxies from \citet{Scholtz2023}. Solid points indicate the narrow component fit in radial bins, with yellow, green, and blue corresponding to the central bin and increasing radial annuli. For bins that include a broad component, the points are displayed with hollow markers, with the same marker shapes and color scheme. Clump bins are also included for GOODS-S~40768 and COSMOS~20062. Star markers correspond to integrated galaxy spectra from different instruments: red for MOSFIRE and blue for OSIRIS. The narrow and broad components to the integrated OSIRIS fit of COSMOS~19985 are shown, while the other galaxies are only fit with one component. The MOSFIRE data of all four galaxies are fit with two components. Arrows indicate 3$\sigma$ upper or lower limits. 
    \label{fig:BPT_diagrams}}
\end{figure*}

The BPT diagram in Fig.~\ref{fig:BPT_diagrams} display the \oiiihb{} versus \niiha{} flux ratios for each galaxy, with solid points representing the narrow components of radial bins, and hollow points denoting the broad components if fitted. The color of each point (yellow, green, and blue) corresponds to the central bin and increasing radial annuli. Clump bins are also included for GOODS-S~40768 and COSMOS~20062. Star markers correspond to integrated galaxy spectra from different instruments: red for MOSFIRE and blue for OSIRIS. To reveal the evolution of emission line ratios towards higher redshifts, we include samples of $z\sim0$ galaxies drawn from the Sloan Digital Sky Survey (SDSS) Data release 7 \citep{Abazajian2009} and $z=1.4-2.7$ galaxies from the AURORA survey \citep{Shapley2025}. These samples are plotted in the background of each BPT diagram as a grayscale histogram and as small red dots respectively. We also overplot the median trend of the $z\sim2$ galaxies from the MOSDEF sample parametrized in \citet{Runco2021} to enable direct visual comparison of our galaxies to the parent sample.

For COSMOS~19985 in the top-left panel, we observe a $\sim0.3$ dex decrease in log(\oiiir/\hbeta) with increasing radius, while log(\niir/\halpha) remains relatively constant. The MOSFIRE spectrum aligns well with the innermost radial bin, and the integrated OSIRIS point is similarly located near the central region. All the points lie above the $z\sim0$ star-forming locus traced by the highest-density region of the grayscale histogram, but are roughly consistent with the $z=1.4-2.7$ galaxies displayed as small red dots.

For COSMOS~20062 in the top-right panel, the central bin \Rsig{1} required a two-component fit, which lowered the \hbeta{} flux assigned to the narrow component, resulting in a non-detection in narrow \hbeta{} flux and a $3\sigma$ lower limit for \oiiihb. Regardless, we find no significant radial trend in \oiiihb{} and a minor ($<0.2$ dex) decrease in \niiha{}. Interestingly, the clump bin, which isolates the upper-left region with high \halpha{} surface brightness, exhibits lower \niiha{} but higher \oiiihb{}. 
% We also observe an offset between the MOSFIRE and integrated OSIRIS points, with the MOSFIRE point being closer to the clump and OSIRIS aligned with the other radial bins. 

AEGIS~3668 (bottom-left panel) shows a similar trend to that of COSMOS~19985, with a $\sim0.3$ dex decline in \oiiihb{} and a minor decrease ($<0.2$ dex) in \niiha{} with increasing radius. The MOSFIRE and integrated OSIRIS points are offset from the radial bins, but the \niiha{} upper limit suggests that the integrated OSIRIS spectrum may still lie within the range spanned by the binned points. Like the previous two COSMOS objects, all the radial bins follow the $z=1.4-2.7$ galaxies and lie below the maximum starburst line. 

In GOODS-S~40768 (bottom-right panel), the clumps span a wide range in \niiha{}, showing a distinct behavior from that of the radial bins of the other three galaxies. All the clump bins have relatively high \oiiihb{} ratios of log(\oiiihb) $\sim 0.7$ as compared to what is observed in the other galaxies. Despite the higher \oiiihb{} ratios, all points except for the center bin lie below the maximum starburst line. The MOSFIRE and integrated OSIRIS spectra agree well with the central, brightest clump, which is expected. 

\begin{figure*}[ht!]
    \centering
    \includegraphics[width=0.65\textwidth]{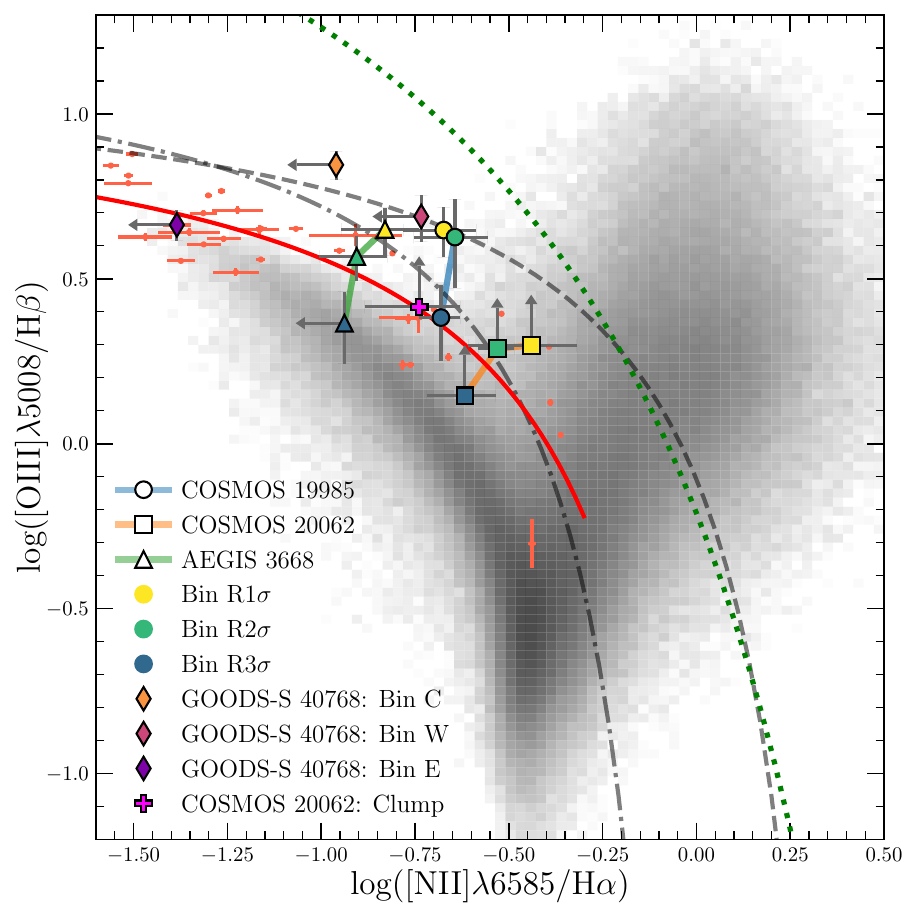} 
    \caption{Summary of BPT emission-line diagnostic diagrams for all galaxies in this study. This figure includes all radial and clump bins from Fig.~\ref{fig:BPT_diagrams}, following the same marker shapes and color scheme. Radial bins for each galaxy are connected by colored lines for clarity. The background grayscale histogram represents the distribution of local SDSS galaxies, small red dots are $z=1.4-2.7$ galaxies from \cite{Shapley2025}, and the red curve denotes the median trend of $z\sim2$ MOSDEF galaxies from \citet{Runco2021}. The dashed, dot-dashed, and dotted curves are from \citet{Kewley2001}, \citet{Kauffmann2003}, and \citet{Scholtz2023} respectively. Arrows indicate 3$\sigma$ upper or lower limits. 
    \label{fig:BPT_diagram_SUMMARY}}
\end{figure*}

Fig.~\ref{fig:BPT_diagram_SUMMARY} shows a compilation of the radial and clump binned points from Fig.~\ref{fig:BPT_diagrams} on a summary BPT diagram. The radial bins for COSMOS~19985, COSMOS~20062, and AEGIS~3668 are connected by colored lines for clarity.

All four galaxies in our sample display the known BPT diagram offset towards higher \niiha{} and \oiiihb{} for $z>1$ galaxies. Moreover, among all our radially binned data points, none lie above the \cite{Kewley2001} maximum starburst threshold, suggesting that inclusion of an AGN is not required to explain the degree of ionization and excitation observed in these galaxies. The \niiha{} ratio of the central bin of GOODS-S~40768 is an upper-limit, and thus can fall below or be consistent with the maximum starburst threshold. Given that the properties of the central region in galaxies are most relevant in the search for AGNs, the fact that none of the central bins exceed the typical AGN identification threshold of \niiha{} $=0.5$ (or log(\niiha{}) $=-0.3$), as well as the lack of drastic differences in the line ratios of such central bins compared to bins at larger radii, suggest that there is no compelling evidence for AGN activity across the galaxies in any of the galaxies in our sample.

The central regions of galaxies are the most relevant for identifying AGNs. In our sample, none of the central bins exceed the typical AGN threshold of \niiha{} $= 0.5$ (or log(\niiha{}) $= -0.3$). Additionally, the line ratios in these central bins are not significantly different from those at larger radii. Together, these findings suggest there is no compelling evidence for AGN activity in any of the galaxies we studied.

In general, we observe a decrease in \oiii/\hbeta{} with increasing radius, while \nii/\halpha{} shows little to no significant variation, with at most a slight decrease. In particular, while the emission line ratios of each galaxy as a whole follow the general shape of the star-forming locus, the variation within individual galaxies appears nearly orthogonal to this trend, suggesting that a decrease in \nii/\halpha{} does not correspond to an increase in \oiii/\hbeta{} within a galaxy.

%----------------------------------------------------%
\subsection{Radial variations in \niiha{} ratios} \label{subsec:radial_niiha}

From the Voronoi-binned K-band data, we obtained the \niiha{} flux ratios for the fitted line profiles in COSMOS~19985, COSMOS~20062, and AEGIS~3668. GOODS-S~40768 was omitted from this analysis due to its unusual morphology and low S/N. As mentioned in Section~\ref{subsubsec:vorbin}, the Voronoi-binning strategy recovers more spatial information than the radial bins used in the BPT diagram analysis, enabling us to more robustly examine spatial variations of emission line ratios.

\begin{figure*}[ht!]
    \centering
    \includegraphics[width=\textwidth]{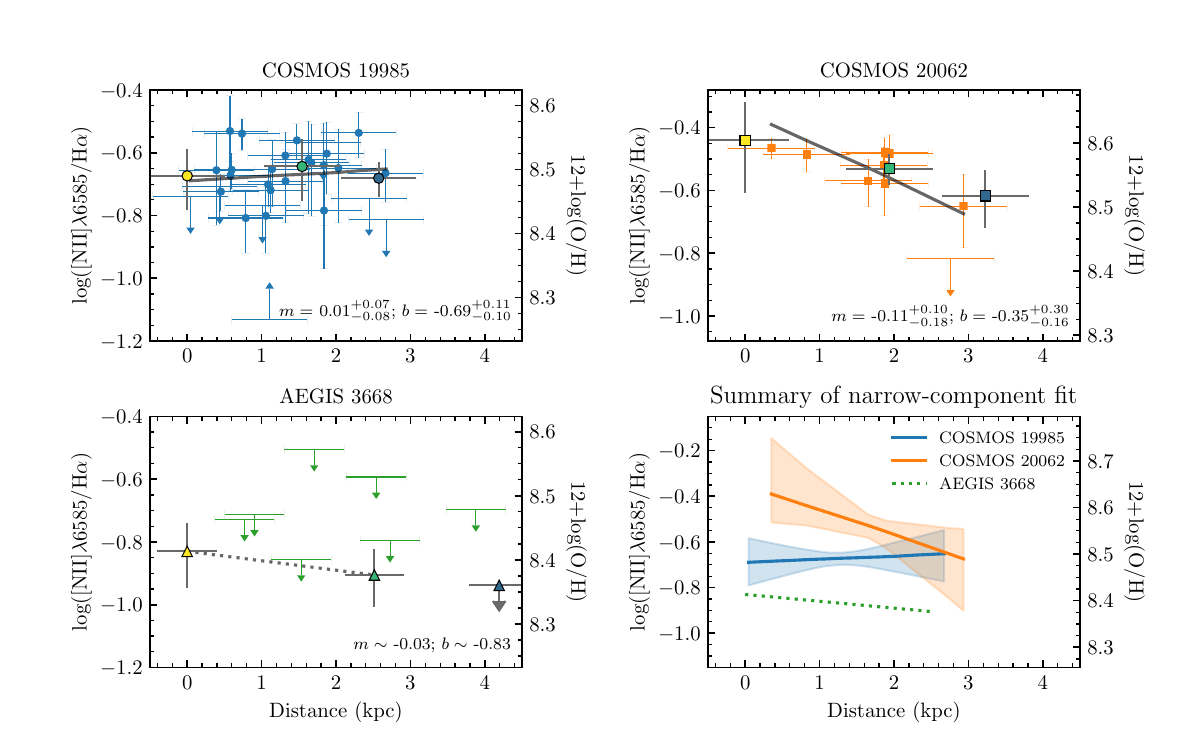} 
    \caption{\niiha{} flux ratio vs. distance for Voronoi-binned fits of COSMOS~19985 (top-left), COSMOS~20062 (top-right), and AEGIS~3668 (bottom-left). The bin distances are from the luminosity-weighted center of each bin to the center of the galaxy ellipse fit. Small scatter points represent the narrow component and arrows indicate 3$\sigma$ upper limits. Points from radial bins are also shown, following the marker shapes and color schemes in Fig.~\ref{fig:BPT_diagrams}. The solid lines show the median linear fits to the scatter points, with the bottom-right panel overlaying all the fitted lines and their 16th–84th percentile error regions. Median slopes and intercepts, along with errors, are reported at the bottom of the COSMOS~19985 and COSMOS~20062 panels. The values quoted for AEGIS~3668 parametrize the dotted line connecting the two detected radial points, and thus should be interpreted with caution.}
    \label{fig:radial_niiha}
\end{figure*}

Fig.~\ref{fig:radial_niiha} presents the \niiha{} ratios in each bin as a function of the distance from the luminosity-weighted center of each of the Voronoi bins to the center of the galaxy ellipse fit. To facilitate interpretation, we also show the conversion of \niiha{} to gas-phase metallicity for high-redshift galaxies, using the relation 
\begin{equation}
     12 + \log(\mathrm{O/H}) = 8.82 + 0.49 \times \log(\mathrm{[NII]}\lambda6585/\mathrm{H}\alpha)
\end{equation}
from \cite{Bian2018}. Though this relation is calibrated for galaxies with metallicities spanning $7.8 < 12+\log(\mathrm{O/H}) < 8.4$ and our analysis requires extrapolation up to $12+\log(\mathrm{O/H}) \sim 8.6$, the uncertainties from extending the relation slightly above the metallicity range should be small in comparison to those resulting from $z\sim0$ calibrations. Furthermore, as the analysis mainly focuses on metallicity gradients, the absolute value calculated from the relation matters less than the relative differences between data points. In Voronoi bins with two-component fits, we apply the strong-line calibration only to the narrow component, which is most representative of H II region emission. We exclude the broad component from this analysis as they may trace outflows or potential weak AGN activity.

In Fig.~\ref{fig:radial_niiha}, the small solid points represent the narrow component and arrows indicate 3$\sigma$ upper limits. Points from radial bins are also shown with larger markers. The yellow, green, and blue markers represent the \Rsig{1}, \Rsig{2}, and \Rsig{3} bins, following the marker shapes and color schemes in Fig.~\ref{fig:BPT_diagrams}.

We performed a linear fit using the Bayesian regression method implemented in \texttt{linmix} \citep{Kelly2007}, which accounts for non-detections through an MCMC-based approach. The plotted lines in the panels in Fig.~\ref{fig:radial_niiha} show the median fit to the narrow component points, with the bottom-right panel displaying the fits along with the corresponding 16th-84th percentile error regions. The median slope $m$ and intercept $b$ of the fit, along with their standard deviations, are reported at the bottom of the panels. In AEGIS~3668, due to a high number of non-detections, we simply connected the two detected points with a dotted line and fit errors are not included, thus any interpretation should be made with caution. 

In COSMOS~19985 (top-left panel), the \niiha{} ratios exhibit a slight positive radial gradient of $\nabla\log$(\niiha{})$=0.01^{+0.07}_{-0.08}$~dex~kpc\sups{-1}, which translates to a metallicity gradient of $\nabla\log(\mathrm{O/H}) = 0.007^{+0.036}_{-0.041}$~dex~kpc\sups{-1}. However, given the large uncertainty, there is no statistically significant preference for a positive or negative trend. The mean \niiha{} value across detected bins is log(\niiha{}) $=-0.66\pm0.08$, corresponding to a metallicity of $8.49\pm0.04$. 
As mentioned in Section~\ref{subsubsec:vorbin}, the bins in the central region of the system are smaller than the PSF, which may cause measurements to be spatially correlated. However, these bins are confined to the center of the galaxy and span only a small fraction of its full radial extent. The inferred gradient is therefore primarily constrained by data points at the outskirts where PSF effects are less significant due to the large bin sizes. Moreover, spatially correlated measurements would tend to artificially wash out intrinsic gradients. The fact that the \niiha{} profiles remain flat at large radii suggests that the overall flat gradients are likely physical, rather than an artifact driven by the central bins.

In COSMOS~20062 (top-right panel), the ratios suggest a negative gradient of $\nabla\log$(\niiha{})$=-0.11^{+0.10}_{-0.18}$~dex~kpc\sups{-1} or $\nabla\log(\mathrm{O/H}) = -0.054^{+0.047}_{-0.086}$~dex~kpc\sups{-1}, but the large uncertainty once again renders the slope insignificant. This apparent trend is likely skewed by the low point at $\sim 2.8$~kpc, corresponding to Voronoi bin 4 (see Fig.~\ref{fig:vorbin_fits/co20062_vorbin_final_fits}), where the \niir{} line flux is particularly weak. The non-detection at the largest radius corresponds to the bin containing the clump described in Section~\ref{subsubsec:radial_binning}, where the large fitting uncertainties drive down the S/N. The mean \niiha{} ratio is log(\niiha{}) $=-0.55\pm0.12$, corresponding to a metallicity of $8.55\pm0.06$.

All the bins in AEGIS~3668 (bottom-left panel) were fitted by a one-component model. However, due to the low S/N of the spectra and weak \nii{}, all \niiha{} ratios from the Voronoi bins were non-detections. Despite the high fraction of non-detections, the two detected points from the radial bins differ by about 0.05 dex over a radial distance of nearly 2.5 kpc, suggesting that there is no significant metallicity gradient. We connected the two detections to infer $\nabla\log$(\niiha{})$\sim-0.03$~dex~kpc\sups{-1} or $\nabla\log(\mathrm{O/H})\sim -0.01$~dex~kpc\sups{-1}. 

The \niiha{} flux ratios measured from the radially-binned spectra also follow those from the Voronoi-binned spectra very closely. Particularly, for COSMOS~20062, points from the radial bins trace the main sequence of Voronoi-binned points, showing that the low point at $\sim 2.8$~kpc is an outlier as compared to the general metallicity trend of the rest of the galaxy. The metallicity gradient measured for COSMOS~20062 is thus driven more negative by the low \niiha{} flux ratio of Voronoi bin 4. 

All in all, the \niiha{} ratios do not show clear positive or negative radial trends across the galaxies in our sample, consistent with the lack of strong radial variation seen in the BPT diagrams in Section~\ref{subsec:BPT}. Across all three galaxies, the \niiha{} ratios at the center-most regions remain below log(\niiha) $\sim -0.3$ (or \niiha{} $\sim 0.5$), a commonly used threshold for distinguishing AGN ionization from star formation \citep[e.g.,][]{Shapley2015}. The absence of higher \niiha{} ratios at small radii indicates that the central regions are not strongly influenced by AGN ionization, and are, instead, more consistent with excitation by star formation.

%----------------------------------------------------%
\subsection{Radial variations in (\oiiihb{})/(\niiha{}) ratios} \label{subsec:radial_o3n2}

Similar to the analysis in Section~\ref{subsec:radial_niiha}, the (\oiiihb{})/(\niiha{}) (O3N2) ratio serves as a metallicity indicator. From the radially-binned H- and K-band data, we obtained O3N2 ratios for the fitted line profiles in COSMOS~19985, COSMOS~20062, and AEGIS~3668 for metallicity gradient analysis. 

\begin{figure*}[ht!]
    \centering
    \includegraphics[width=\textwidth]{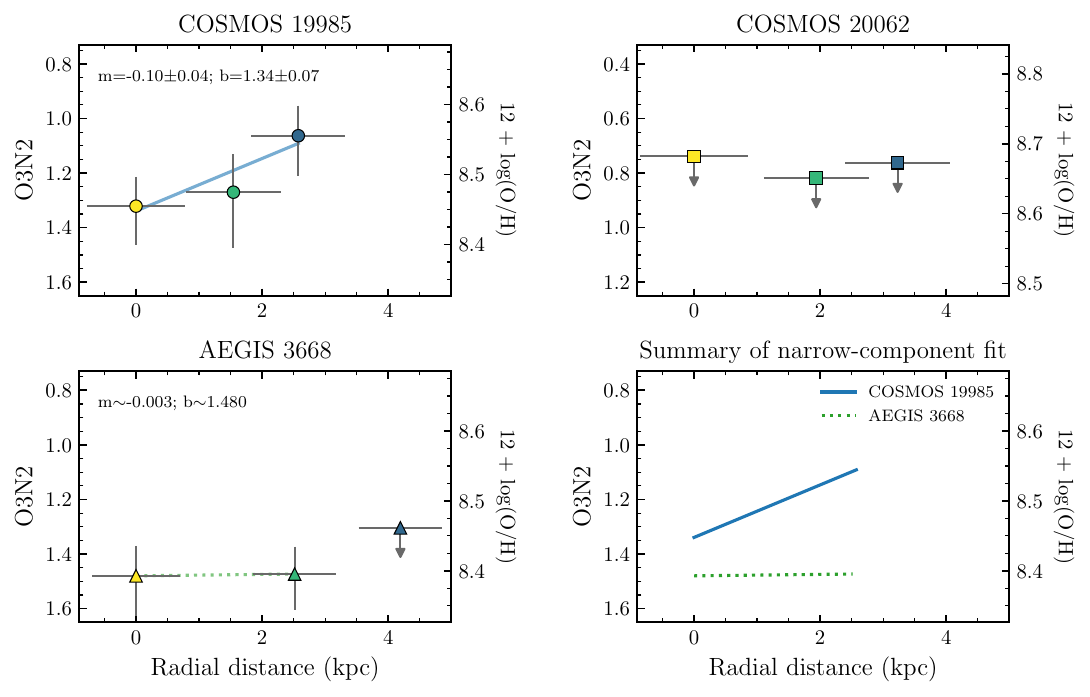} 
    \caption{(\oiiihb{})/(\niiha{}) O3N2 ratios vs. distance for radially-binned fits of COSMOS~19985 (top-left), COSMOS~20062 (top-right), and AEGIS~3668 (bottom-left). The bin distances are from the mid-point of each radial annuli to the center of the galaxy ellipse fit. Solid points represent the narrow component and arrows indicate 3$\sigma$ upper limits. The marker shapes and color schemes follow those in Fig.~\ref{fig:BPT_diagrams}. The dotted lines show the linear fits to the narrow component, with the bottom-right panel overlaying all the fitted lines. Given the small number of bins and detections, the fitted slope and intercept values shown in each panel should be interpreted with caution.}
    \label{fig:radial_o3n2}
\end{figure*}

Fig.~\ref{fig:radial_o3n2} shows the O3N2 ratios in the radial bins as a function of the distance from the bins to the center of the galaxy ellipse fit. The distance is defined as the center of the galaxy ellipse fit for \Rsig{1}, and 1.5 and 2.5 times the semi-major axis of the ellipse fit for \Rsig{2} and \Rsig{3}. The conversion of O3N2 to gas-phase metallicity for high-redshift galaxies, using the relation 
\begin{equation}
     12 + \log(\mathrm{O/H}) = 8.97 - 0.39 \times \log(\mathrm{O3N2})
\end{equation}
from \cite{Bian2018}. Again, we note that the highest metallicity in our sample exceeds the upper limit of $12 + \log(\mathrm{O/H})=8.4$ of this calibration by 0.2~dex. Though extrapolation is needed, we determine that analysis performed based on this relation is still more robust than that based on $z\sim0$ calibrations. The data points follow the marker shapes and color schemes in Fig.~\ref{fig:BPT_diagrams}, where solid points denote the narrow component. Arrows indicate 3$\sigma$ upper limits. The O3N2 axes are inverted to reflect the inverse relationship between O3N2 and metallicity.

For COSMOS~19985, we fit the three data points with a linear model using a least-squares approach. Given that only two radial bins in AEGIS~3668 have detections in O3N2, we simply connected the two detections with a straight line. All radial bins in COSMOS~20062 are non-detections and thus are not fit with a gradient. These linear fits are overplotted as dotted lines in each panel. Due to the small number of bins and detections, the fitted slope $m$ and intercept $b$ values shown in each panel should be interpreted with caution. 

In COSMOS~19985 (top-left panel), the O3N2 ratios show a negative trend of $\nabla\log(\mathrm{O3N2})\sim-0.10\pm0.04$~dex~kpc\sups{-1}, which gives a slightly positive metalicity gradient of $\nabla\log(\mathrm{O/H})\sim0.04\pm0.01$~dex~kpc\sups{-1}, similar to that derived from the \niiha{} ratio in Section~\ref{subsec:radial_niiha}. The mean O3N2 value across detected narrow component points is $\log(\mathrm{O3N2}) = 1.22\pm0.11$, which corresponds to a metallicity of $8.49\pm0.04$ and is in agreement with \niiha{}-derived values.

For AEGIS~3668, the O3N2 gradient and metallicity gradient from the two detections in \Rsig{1} and \Rsig{2} are flat, with $\nabla\log(\mathrm{O3N2})\sim-0.003$~dex~kpc\sups{-1} or $\nabla\log(\mathrm{O/H})\sim0.001$~dex~kpc\sups{-1}. The mean O3N2 ratio among the two detected points is $\log(\mathrm{O3N2}) \sim 1.48$, corresponding to a metallicity of $\sim 8.39$ which is consistent with the metallicity calculated with the \niiha{} diagnostic.

All in all, the O3N2-based gas-phase metallicities and metallicity gradients corroborate those derived using the \niiha{} calibrations.

%%%%%%%%%%%%%%%%%%%%%%%%%%%%%%%%%%%%%%%%%%%%%%%%%%%%%%%%%%%%%%%%%%%%%%%%%%%%%%%%%%%%%%%%%%%%%%%%%%
%%%%%%%%%%%%%%%%%%%%%%%%%%%%%%%%%%%%%%%%%%%%%%%%%%%%%%%%%%%%%%%%%%%%%%%%%%%%%%%%%%%%%%%%%%%%%%%%%%
\section{Discussion} \label{sec:discussion}
\subsection{AGN diagnostics} \label{subsec:AGN}

We present spatially resolved IFU data on four galaxies at $z\sim2$, previously observed with integrated spectroscopy as part of the MOSDEF survey. The classification of these galaxies as star-forming or AGN based on emission line ratios is complicated by the known offset of the high-redshift star-forming locus on the BPT diagram, which shifts towards higher \oiiihb{} and \niiha{} values relative to the sequence defined by local star-forming galaxies \citep[e.g.,][]{Steidel2014, Shapley2015}. As discussed previously, among many possible explanations, it appears that this BPT offset is driven by a harder ionizing spectrum at fixed nebular metallicity for chemically-young, high-redshift galaxies \citep[e.g.,][]{Steidel2016,Cullen2021,Shapley2025}.

The offset of the $z\sim2$ star-forming locus causes many high-redshift star-forming galaxies to occupy the ``composite" region of the BPT diagram based on calibrations at $z\sim0$. In this region, AGN and star-formation contributions are difficult to disentangle. This ambiguity can be seen in our sample, with COSMOS~19985 and COSMOS~20062 being classified as star-forming or as hosting AGN based on their integrated spectra in different studies \citep{Sanders2023, Leung2017}. To make matters even more complicated, emission line ratios obtained from integrated galaxy spectra can differ from those measured from smaller apertures focused on central regions in galaxies that host a weak AGN. \citet{Wright2010} demonstrated this effect for distant-galaxy spectra, measuring a $\sim 0.3$~dex increase in both log(\oiiihb{}) and log(\niiha{}) for a $z=1.6$ galaxy's ``concentrated" spectrum relative the integrated measurements. These discrepancies and uncertainties highlight the need for high-redshift calibrations of emission line diagnostics, as well as spatially-resolved observations to better isolate the ionization mechanisms in the central regions of galaxies where AGN activity is expected to have the strongest influence on line ratios.

Therefore, in Section~\ref{subsec:BPT}, we primarily used two demarcations lines to interpret the nature of the galaxies in our sample: the maximum starburst line from \citet{Kewley2001} that sets a theoretical threshold for ionization from pure star formation, and the empirical AGN threshold based on observations of $z<5$ galaxies from \citet{Scholtz2023}. In \citet{Scholtz2023}, high-redshift Type 2 AGN are seen to occupy similar spaces as star-forming galaxies, so the latter boundary serves as a conservative limit above which no star-forming galaxies are found. We find that all but two of our data points fall below the maximum starburst line and that all points lie well below the \citet{Scholtz2023} limit. 

We observe an increase in \oiiihb{} ratio in the center-most bin \Rsig{1} in COSMOS~19985 as compared to the value obtained from the integrated OSIRIS spectrum. Similarly, we see an increase in both \oiiihb{} and \niiha{} for AEGIS~3668. This trend reflects the effect of aperture size discussed above and in \citet{Wright2010}, where the contributions of an embedded weak AGN could be drowned out by the light of a predominantly star-forming galaxy in an integrated spectrum of the galaxy. While smaller apertures focused on the central regions of a galaxy, or bin \Rsig{1} in our analysis, can help isolate the contribution of an AGN if present, our data show that none of the central bins exhibit line ratios elevated beyond the maximum starburst line. Combined with the fact that our data points scatter around those from $z=1.4-2.7$ galaxies from \citet{Shapley2025} and that the centermost, finer Voronoi bins have \niiha{} $<0.5$, there is no compelling evidence for AGN activity in our sample, and the AGN scenario is disfavored.

\subsection{Notable deviations and unusual trends in the BPT diagram} \label{subsec:special_cases}

As mentioned in Section~\ref{subsec:BPT}, the center radial or clump bins of COSMOS~19985 and GOODS-S~40768 lie on or slightly above the maximum starburst threshold. In COSMOS~19985, the central spaxel within radial bin \Rsig{1} as probed by Voronoi bin 0 required a broad \halpha{} component with FWHM velocity $\sim1200$~km~s\sups{-1} to fit the emission line profile (see Section~\ref{subsubsec:spec_co19985}). However, the broad \halpha{} flux is barely detected, with S/N just above 3, making it unlikely that we would detect a corresponding broad \niir{} component even if present. Additionally, the high star formation rate (SFR) surface density of 18.5$\pm$0.7~M$_{\odot}$~yr\sups{-1}~kpc\sups{-2} measured by \citet{Sanders2023} significantly exceeds the 0.1~M$_{\odot}$yr\sups{-1}~kpc\sups{-2} requirement for star formation-driven outflows. Combined with the observed \niiha{}~$\ll 0.5$ in the central spaxel, these observations point towards complex outflow kinematics rather than AGN activity in COSMOS~19985.

GOODS-S~40768 also exhibits an anomalous feature, with the center clump bin showing the most significant offset above the maximum starburst line among our sample (see Fig.~\ref{fig:BPT_diagram_SUMMARY}). Though strong central AGN activity is unlikely, further observations of additional emission lines (e.g. [O\thinspace{\sc i}], [Si\thinspace{\sc ii}], [Ne\thinspace{\sc iii}], [O\thinspace{\sc ii}]) are needed to better characterize the ionization conditions in each of the clumps and determine whether the deviation in the center clump is due to unusual ISM conditions, a hidden AGN contribution, shocks from accreting gas, or other mechanisms.

Beyond individual points above the maximum starburst line, we also observe a notable internal trend in our galaxies with radial binning where higher \oiiihb{} ratios do not correspond to lower \niiha{}. This pattern diverges from typical expectations for star-forming galaxies and suggests the presence of additional physical processes influencing the observed line ratios. However, further data are needed to confirm and interpret this trend.

\subsection{Radial metallicity gradients} \label{subsec:rad_metal_grad}

In Sections~\ref{subsec:radial_niiha} and \ref{subsec:radial_o3n2}, we investigated radial trends in \niiha{} and O3N2 ratios measured from Voronoi-binned and radially-binned spectra to infer the gas-phase metallicity gradients for our galaxies. The mean \niiha{} ratios measured for COSMOS~19985 and COSMOS~20062 are consistent with those reported in \citet{Sanders2023}, though our inferred metallicity for COSMOS~19985 is 0.5~dex higher than the direct metallicity obtained in their work. While the absolute metallicity scaling inferred from \niiha{} may be subject to systematic offsets at high redshifts, our analysis focuses on the relative gradient rather than the absolute values.

Among the three galaxies on which we performed Voronoi binning, we find our targets exhibit mostly flat radial gradients in gas-phase metallicity, with COSMOS~19985 and AEGIS~3668 showing gradients within $\nabla\log(\mathrm{O/H}) = 0.01$~dex~kpc\sups{-1} positive and negative from flat respectively. The steepest gradient of $\nabla\log(\mathrm{O/H}) = -0.054$~dex~kpc\sups{-1} observed in COSMOS~20062 has a large uncertainty that can shift the slope to be nearly consistent with zero slope. The metallicity gradients calculated from the O3N2 diagnostic with the fewer, but higher S/N radially-binned data, also corroborate the flat trends revealed by the \niiha{} ratio analysis.

Flat gas-phase metallicity gradients have been widely reported for star-forming galaxies at $z\sim2$ and are commonly described by efficient radial mixing driven by supernova feedback and galactic outflows \citep[e.g.,][]{Gibson2013,Ma2017,Curti2020,FS2018}. A similar prevalence of flat gradients has been observed in slitless HST grism spectroscopy, where 54 out of 76 galaxies at $z\sim2$ show metallicity gradients consistent with flat within $2\sigma$ uncertainties \citep{Wang2020}. However, there is a significant diversity of gradients in the literature, with values ranging from $-0.2$ to $+0.2$~dex~kpc\sups{-1} \citep{Wang2020}. Some studies have found steep negative gradients in lensed galaxies at $z\sim2$, suggesting inside-out growth \citep{Jones2010,Jones2013}. Others report a high fraction of positive (or inverted) gradients in overdense regions as compared to the fractions found in blank fields at $z\sim2$, possibly driven by ``cold-mode" accretion of metal-poor gas into central regions of galaxies \citep{Wang2019,Li2022}. The large scatter in observed metallicity gradients at $z\sim2$ may reflect the diversity of physical mechanisms shaping metallicity distributions at this epoch.

Cosmological simulations have similarly found a wide range of metallicity gradients at high redshift, depending on the strength of stellar feedback and the timescales of metal redistribution. High-resolution simulations with strong stellar feedback tend to predict flatter gradients at high-redshift, as outflows efficiently mix metals throughout the ISM \citep[e.g.,][]{Ma2017, Tissera2019}. In contrast, lower-resolution simulations with weaker subgrid feedback prescriptions tend to produce more negative gradients, as metals remain more centrally concentrated \citep[e.g.,][]{Gibson2013, Hemler2021}. Additionally, \citet{Ma2017} demonstrated that, at high redshift, short-timescale variations in metallicity gradients can occur due to episodic starburst activity. During a starburst phase, infalling metal-poor gas can temporarily steepen a negative gradient, while subsequent outflows redistribute metals, flattening the gradient. Over time, metal recycling via galactic fountains can re-establish a negative gradient. This effect is particularly pronounced in low-mass, high-SFR galaxies where stellar feedback is most efficient. Given the high SFR of our galaxies and relatively flat metallicity gradients, our observations are consistent with a scenario with significant radial mixing and redistribution of metals by star formation-driven outflows.

Furthermore, correlations between metallicity gradients and global galaxy properties also vary across studies. In overdense environments, such as galaxy clusters or protoclusters, metallicity gradients tend to be flat or inverted, with evidence of anticorrelation between metallicity gradients and global metallicities \citep{Li2022}. \citet{Curti2020} also found that in galaxies with significantly inverted (positive) metallicity gradients, regions with higher SFR surface density tend to have weaker (less positive) metallicity gradients, while regions with lower SFR surface density tend to have stronger (more positive) metallicity gradients. This indicates an anticorrelation between metallicity gradients and SFR surface density on local scales, suggesting episodes of pristine gas accretion or strong radial flows. On the other hand, field galaxies tend to have flat or slightly negative gradients, with no significant correlation between metallicity gradients and global galaxy properties, kinematics, or AGN activity \citep{FS2018}. Since our targets are field galaxies, their flat gradients may reflect the lower environmental influence expected in field regions. Along with the detection of gas flow components in COSMOS~19985 and COSMOS~20062, our results potentially highlight the role of internal processes such as outflows and radial mixing over global environmental effects in the flattening of gas-phase metallicity gradients.

Another factor that impacts measured metallicity gradients is insufficient spatial sampling. Low spatial resolution can artificially flatten metallicity gradients \citep[e.g.,][]{Yuan2013, Acharyya2020}. While some of our inner Voronoi bins cover sub-kpc regions, most outer bins span areas larger than 1~kpc$^2$, which can contribute to the flattening of radial trends. This effect is most pronounced in COSMOS~20062 and AEGIS~3668, where coarse Voronoi binning was necessary to achieve sufficient S/N. However, nearly all Voronoi bins in COSMOS~19985 have sub-kpc spatial resolution, ensuring that any intrinsic metallicity gradient would be clearly detected. The fact that COSMOS~19985 still exhibits a flat metallicity gradient strongly indicates that this result is not an artifact of limited spatial resolution, but rather a true feature of the galaxy.

%%%%%%%%%%%%%%%%%%%%%%%%%%%%%%%%%%%%%%%%%%%%%%%%%%%%%%%%%%%%%%%%%%%%%%%%%%%%%%%%%%%%%%%%%%%%%%%%%%
%%%%%%%%%%%%%%%%%%%%%%%%%%%%%%%%%%%%%%%%%%%%%%%%%%%%%%%%%%%%%%%%%%%%%%%%%%%%%%%%%%%%%%%%%%%%%%%%%%
\section{Conclusion} \label{sec:conclusion}

High-redshift galaxies have different ISM physical conditions from those found in the local universe. While these differences limit the reliability of locally calibrated emission line ratio diagnostics, by studying the variations, we can gain insight into the mechanisms that drive galaxy evolution and shape the ISM during this epoch of peak star formation.

In this work, we conducted a spatially resolved analysis of four galaxies at $z\sim2$ selected from the MOSDEF survey for their potentially composite nature from BPT diagnostics and possible evidence of strong gas flow. Our analysis of IFU data revealed that the \niiha{} and \oiiihb{} ratios in the central $\sim1$~kpc region of our galaxies are not high enough to provide compelling evidence for AGN activity, and instead supports the interpretation that these galaxies are undergoing intense star formation, capable of driving strong outflows. Furthermore, the gas-phase metallicity gradients, as probed by \niiha{} ratios, appear to be flat. This result aligns with previous findings at similar redshifts and suggests that efficient radial mixing shapes the distribution of metals within the ISM during cosmic noon.

Beyond these global trends, we identified intriguing features within individual galaxies, including radial BPT variations that deviate from the expected star-forming sequence and localized regions that lie slightly above the maximum starburst line. In order to fully disentangle the ionizing mechanisms and physical processes responsible for the deviations above the theoretical maximum of ionization from pure star-formation, we require data on additional emission lines such as [O\thinspace{\sc i}], [Si\thinspace{\sc ii}], [Ne\thinspace{\sc iii}], and [O\thinspace{\sc ii}]. Flux ratios measured from these emission lines will allow us to probe shocks, constrain ionization parameters, and reveal the ionizing spectra, all of which would provide a clearer picture of the true nature of these galaxies. 

All in all, expanding the sample of spatially resolved studies at high redshift will be crucial for building a more complete picture of the ISM conditions and feedback processes driving galaxy evolution during cosmic noon. Observation with JWST's NIRSpec IFU enables high S/N detections of faint rest-UV and optical emission lines at high spatial resolution, which provides more detailed mapping of ionization and chemical properties in the ISM. Meanwhile, ALMA observations of emission lines like [C\thinspace{\sc ii}] in the far-IR can probe neutral gas and enable multi-phase analyses of the ISM. A broader, multi-wavelength approach incorporating these datasets will be essential for disentangling the interplay between star formation, feedback, and chemical enrichment in the early universe.

%%%%%%%%%%%%%%%%%%%%%%%%%%%%%%%%%%%%%%%%%%%%%%%%%%%%%%%%%%%%%%%%%%%%%%%%%%%%%%%%%%%%%%%%%%%%%%%%%%
%%%%%%%%%%%%%%%%%%%%%%%%%%%%%%%%%%%%%%%%%%%%%%%%%%%%%%%%%%%%%%%%%%%%%%%%%%%%%%%%%%%%%%%%%%%%%%%%%%
\section{Acknowledgments}
We acknowledge support from NSF AAG grants AST-2009313 and AST-2307622. 
Some of the data presented herein were obtained at Keck Observatory, which is a private 501(c)3 non-profit organization operated as a scientific partnership among the California Institute of Technology, the University of California, and the National Aeronautics and Space Administration. The Observatory was made possible by the generous financial support of the W. M. Keck Foundation.
The authors wish to recognize and acknowledge the very significant cultural role and reverence that the summit of Maunakea has always had within the Native Hawaiian community. We are most fortunate to have the opportunity to conduct observations from this mountain. 
This work is based on observations taken by the CANDELS Multi-Cycle Treasury Program with the NASA/ESA HST, which is operated by the Association of Universities for Research in Astronomy, Inc., under NASA contract NAS5-26555. Some of the data presented in this paper were obtained from the Mikulski Archive for Space Telescopes (MAST) at the Space Telescope Science Institute. The specific observations analyzed can be accessed via \dataset[https://doi.org/10.17909/T94S3X]{https://doi.org/10.17909/T94S3X}. STScI is operated by the Association of Universities for Research in Astronomy, Inc., under NASA contract NAS5–26555. Support to MAST for these data is provided by the NASA Office of Space Science via grant NAG5–7584 and by other grants and contracts.
\facility{Keck:I (OSIRIS)}

\bibliography{main}{}
\bibliographystyle{aasjournal}

\appendix
\section{Spectra and fitted profiles}\label{app_sec:spectra}

\subsection{MOSFIRE and OSIRIS integrated spectra}\label{app_subsec:mos_int}

\begin{figure*}[ht!]
    \centering
    \includegraphics[width=0.8\textwidth]{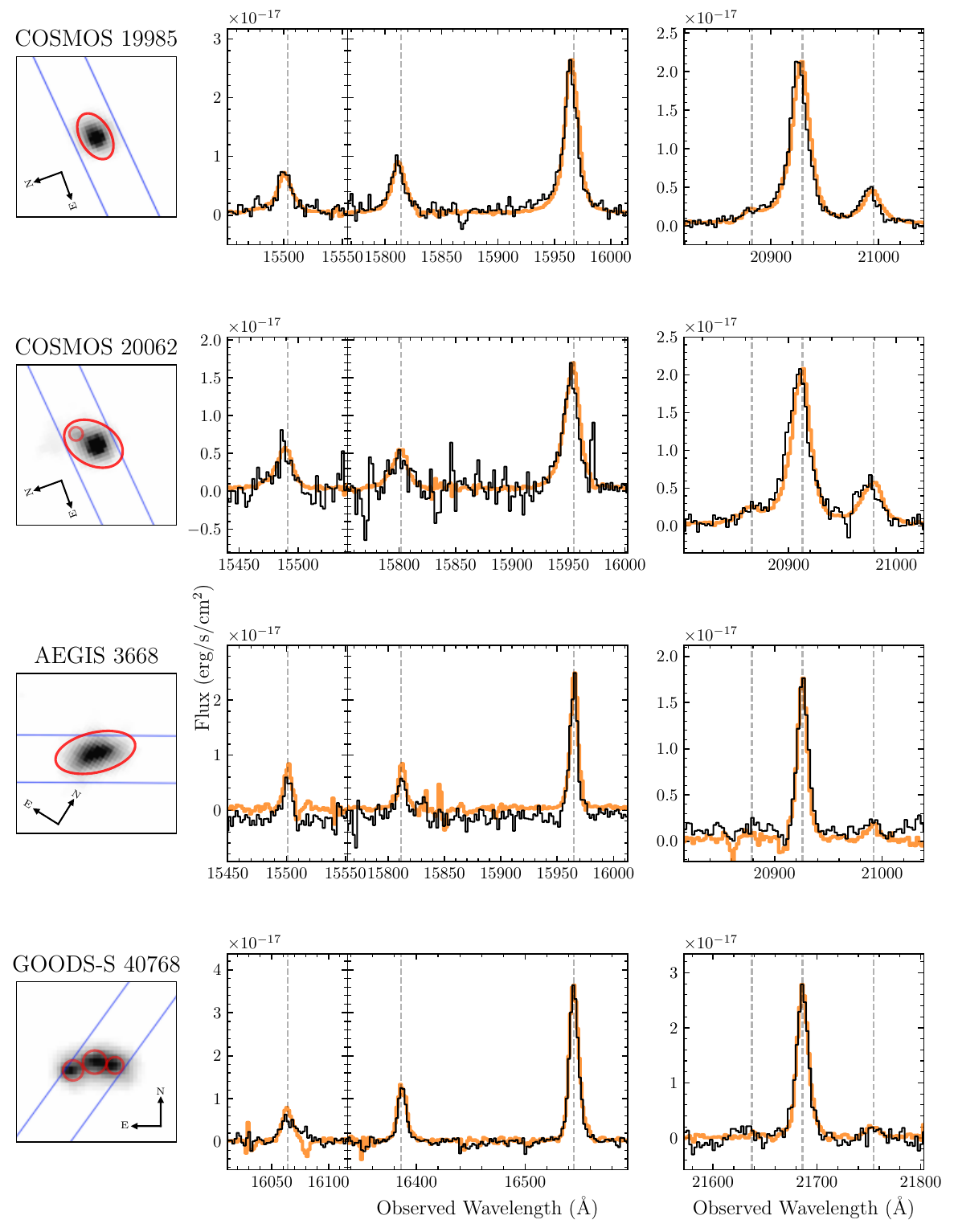} 
    \caption{OSIRIS and MOSFIRE integrated spectra. The middle column shows \hbeta{} and \oiii{} from the H band, while the right column shows \nii{} and \halpha{} from the K band. The left column shows 2.4$''$ HST F160W cutouts of the galaxies, rotated to match the orientation of the OSIRIS observations. The OSIRIS apertures are shown as the 3$\sigma$ ellipses and clump apertures in red and the MOSFIRE slits are overlaid in blue. The integrated spectra (black curves) are generated by summing the spectra in the spaxels within the 3 spatial-$\sigma$ elliptical galaxy mask for COSMOS~19985, COSMOS~20062, and AEGIS~3668 and the spaxels within the three clump bins in GOODS-S~40768. The orange overlaid curves are the spectra taken by Keck/MOSFIRE. The OSIRIS spectra were normalized to those from MOSFIRE by matching the peak flux values of \oiiir{} for the H band and \halpha{} for the K band.
    \label{fig:mos_vs_int}}
\end{figure*}

In this section of the appendix, we compare the Keck/MOSFIRE and Keck/OSIRIS integrated spectra for our sample, as seen in Fig.~\ref{fig:mos_vs_int}. The left column shows 2.4$''$ HST F160W cutouts of the galaxies from the CANDELS survey \citep{Grogin2011,Koekemoer2011}, rotated to match the orientation of the OSIRIS observations. The OSIRIS apertures are shown as the 3$\sigma$ ellipses and clump apertures in red and the MOSFIRE slits are overlaid in blue. The spectra from MOSFIRE and the integrated spectra from OSIRIS are plotted in orange and black respectively in the right columns. The spectra from the MOSDEF survey have been corrected for slit losses following the procedure described in \citet{Kriek2015}, which accounts for galaxy morphology measured from HST imaging. The OSIRIS spectra have been corrected for the offset as determined by the OH line centroiding routine described in Section~\ref{subsec:reduction}, but retain the systemic redshift determined from the MOSDEF survey. As discussed in Section~\ref{subsec:reduction}, there is an offset in wavelengths between OSIRIS and MOSFIRE measurements, particularly for COSMOS~19985 and COSMOS~20062 where their OSIRIS spectra appear blueshifted with respect to those of MOSFIRE. Here, the integrated OSIRIS spectra, generated by summing spaxels within the 3 spatial-$\sigma$ elliptical mask (or within the three clump bins in GOODS-S~40768), are normalized to the MOSFIRE spectra using the peak flux values of \oiiir{} in the H band and \halpha{} in the K band.

\begin{figure*}[ht!]
    \centering
    \includegraphics[width=0.8\textwidth]{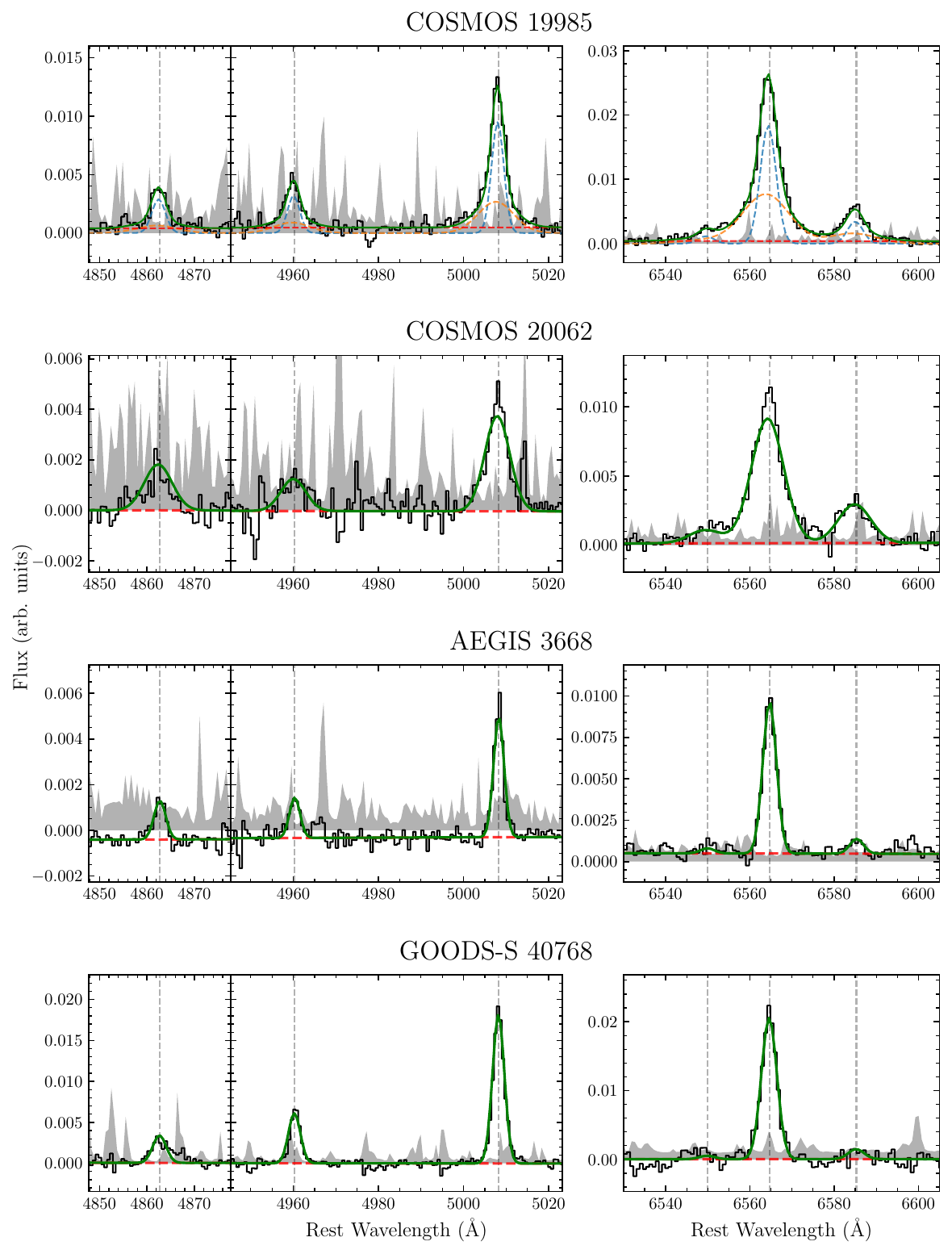} 
    \caption{OSIRIS integrated spectra and fitted profiles. The black curves and gray shaded regions represent the flux and error spectra. The green solid curve represents the total fit, while the red dashed line shows the linear continuum fit around the emission lines. For COSMOS~19985, which prefers a two-component model fit, the blue dashed line corresponds to the narrow component, and the orange line represents the broad component. Locations of strong emission lines are overlaid as light-gray vertical dotted-lines for reference.
    \label{fig:int_fits}}
\end{figure*}

\begin{figure*}[ht!]
    \centering
    \includegraphics[width=0.9\textwidth]{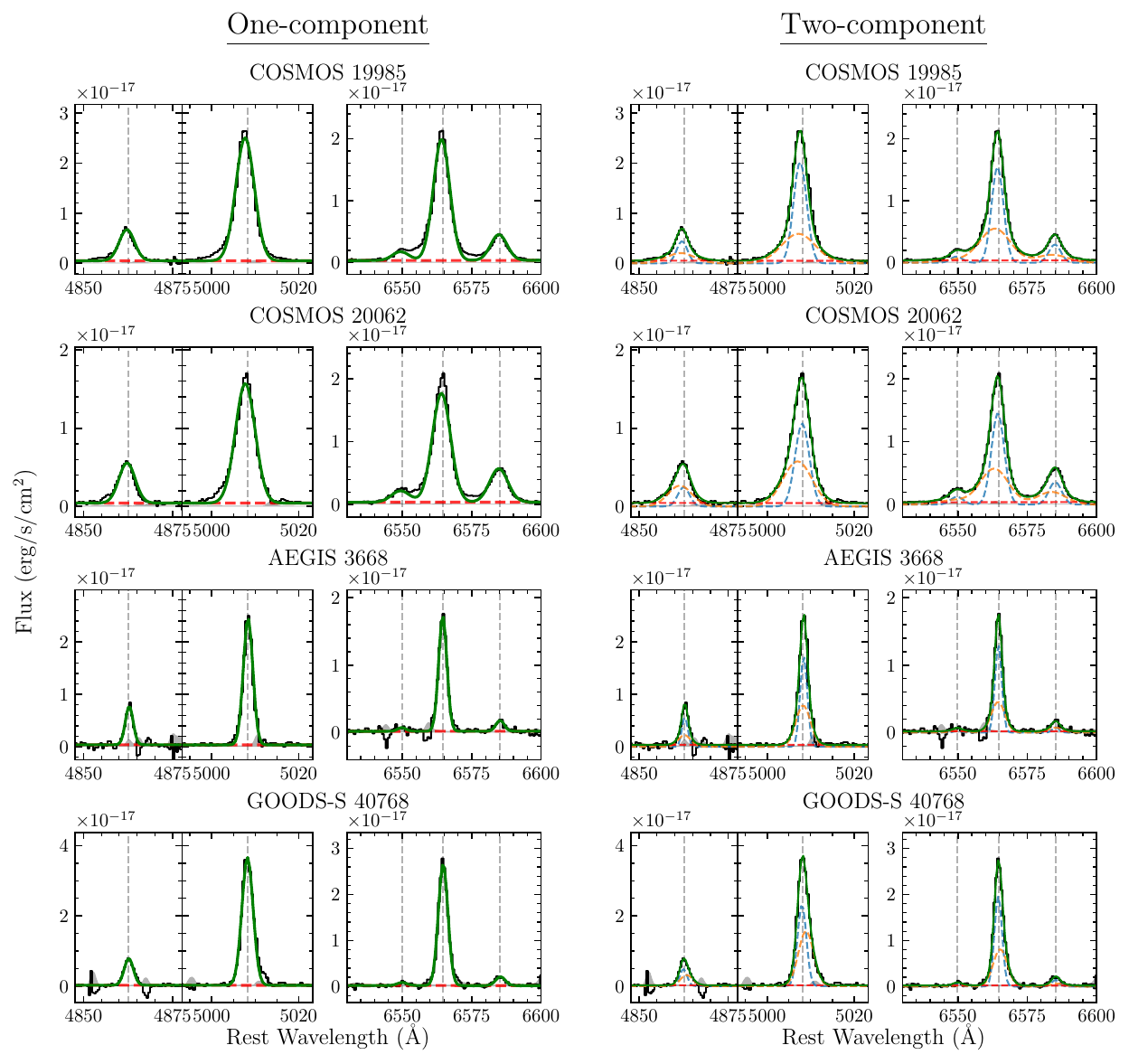} 
    \caption{MOSFIRE spectra and fitted profiles. One-component and two-component fits are shown on the left and right respectively. In each column, the left subcolumn shows \hbeta{} and \oiiir{} from the H band, while the right subcolumn shows \nii{} and \halpha{} from the K band. The black curves and gray shaded regions represent the flux and error spectra. The green solid curve represents the total fit, while the red dashed line shows the linear continuum fit around the emission lines. For the two-component model fit,  the blue dashed line corresponds to the narrow component, and the orange line represents the broad component. Locations of strong emission lines are overlaid as light-gray vertical dotted-lines for reference.
    \label{fig:mos_fits}}
\end{figure*}

Fig.~\ref{fig:int_fits} and \ref{fig:mos_fits} display the OSIRIS integrated spectra and MOSFIRE spectra respectively. Each panel shows the observed spectrum (black curve) along with the error spectrum (gray shaded region). For the model fits, the blue dashed line corresponds to the narrow component, and the orange line represents the broad component where applicable. The total fit is shown in green, with the linear continuum fit overlaid in red. Locations of key emission lines are indicated with vertical light-gray dotted lines for reference. We note that there is a systematic offset in the continuum of AEGIS~3668 in Fig.~\ref{fig:int_fits} that arises from imperfect sky subtraction. However, this offset has minimal impact on our reported flux measurements as it remains within the noise level and is accounted for during the emission line-fitting process where we further subtract a locally-defined linear continuum. Moreover, we do not quote continuum-dependent quantities such as equivalent widths.

For the MOSFIRE spectra , all four galaxies prefer a two-component model fit. However, due to the limited S/N of the OSIRIS integrated spectra, only COSMOS~19985 prefers the two-component model as seen in Fig.~\ref{fig:int_fits}. We adopt the preferred component model of each spectrum individually. This means that COSMOS~19985 is fit with two components in both its integrated OSIRIS and MOSFIRE spectra, while the other three objects have one-component fits for their integrated OSIRIS spectra and two-component fits for their MOSFIRE spectra. We highlight the ambiguity when comparing a one-component fit to either the narrow or broad component of a two-component model, and thus caution against direct comparison between the integrated OSIRIS and MOSFIRE measurements.

% To ensure consistency in comparisons between the MOSFIRE and OSIRIS integrated spectra, the final fitted line profile of the MOSFIRE spectra are set to adopt the preferred component model (i.e. one- or two-components) of the corresponding OSIRIS integrated spectra.
% While a one-component model fit is passable for the MOSFIRE spectra of AEGIS~3668 and GOODS-S~40768, we draw attention to the fact that such a model does not encapsulate the flux of COSMOS~20062 well, especially for \halpha{} in the K band (Fig.~\ref{fig:int_fits}). A detailed comparison of more complex models is presented in Appendix Section~\ref{app_sec:bic}.

\subsection{Radial and clump bins}\label{app_subsec:radial}

\begin{figure*}[ht!]
    \centering
    \includegraphics[width=0.8\textwidth]{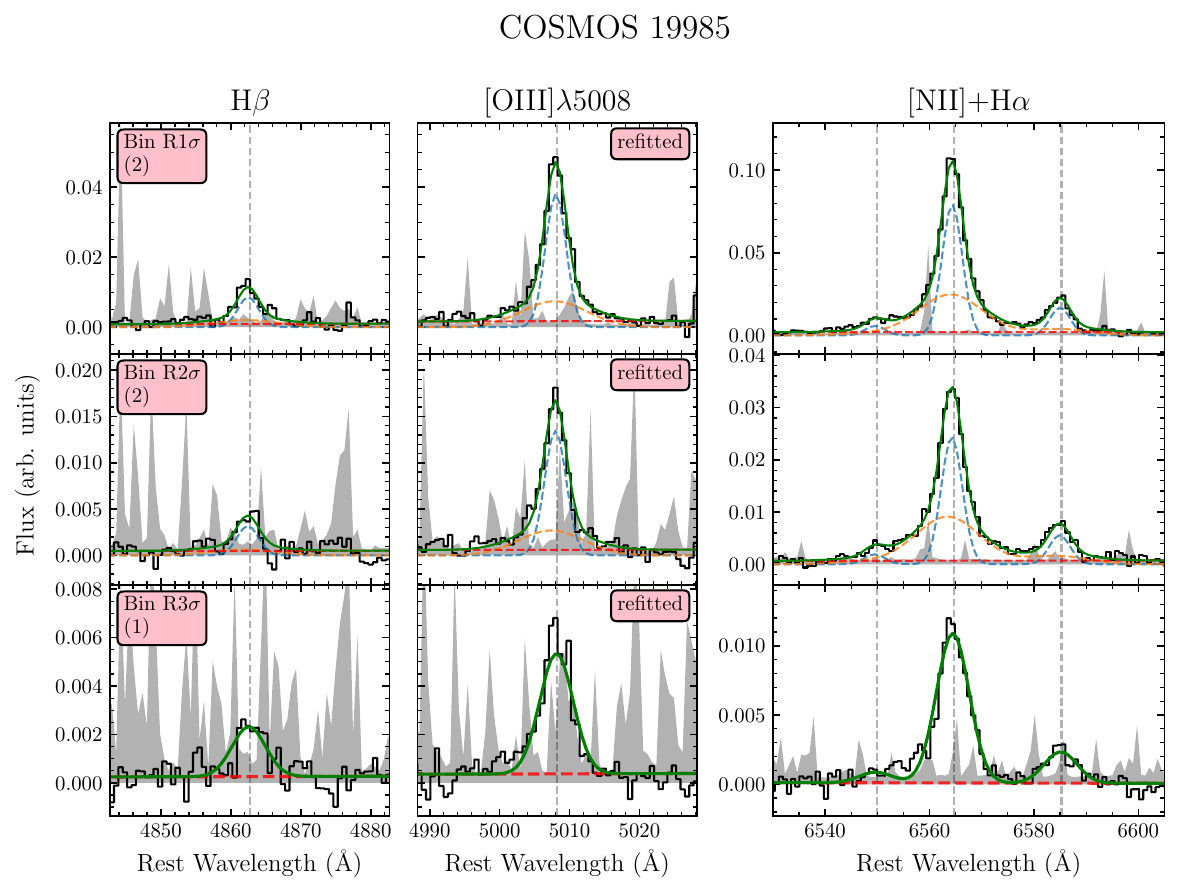} 
    \caption{Radially-binned spectra for COSMOS~19985 and fitted line profiles. The left column shows \hbeta{} and \oiiir{} from the H band and the right column shows the \nii{}+\halpha{} group from the K band. The color scheme follows that of Fig.~\ref{fig:int_fits}. The red text box contains the bin name and whether a one-component model (1) or two-component model (2) is preferred and plotted. A ``refitted" tag is displayed at the top right of the band that underwent refitting as determined by the routine described in Section~\ref{subsubsec:emline_fitting}.
    \label{fig:rad_fits/co19985_fits}}
\end{figure*}

\begin{figure*}[ht!]
    \centering
    \includegraphics[width=0.8\textwidth]{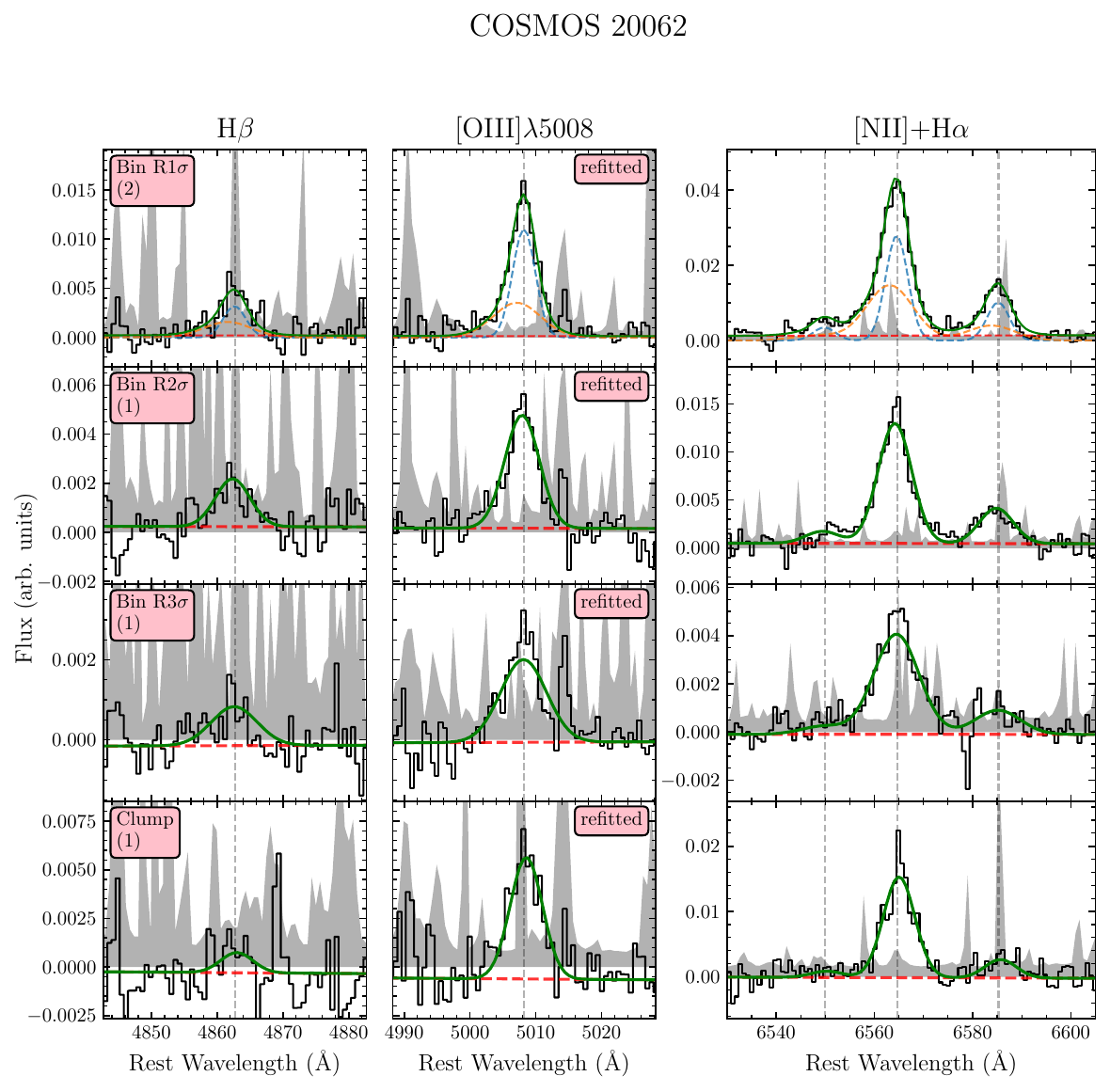} 
    \caption{Radially-binned spectra for COSMOS~20062 and fitted line profiles, similar to Fig.~\ref{fig:rad_fits/co19985_fits}. The last row shows the clump bin shown in Fig.~\ref{fig:flux_bin_maps_co20062}.
    \label{fig:rad_fits/co20062_fits}}
\end{figure*}

\begin{figure*}[ht!]
    \centering
    \includegraphics[width=0.8\textwidth]{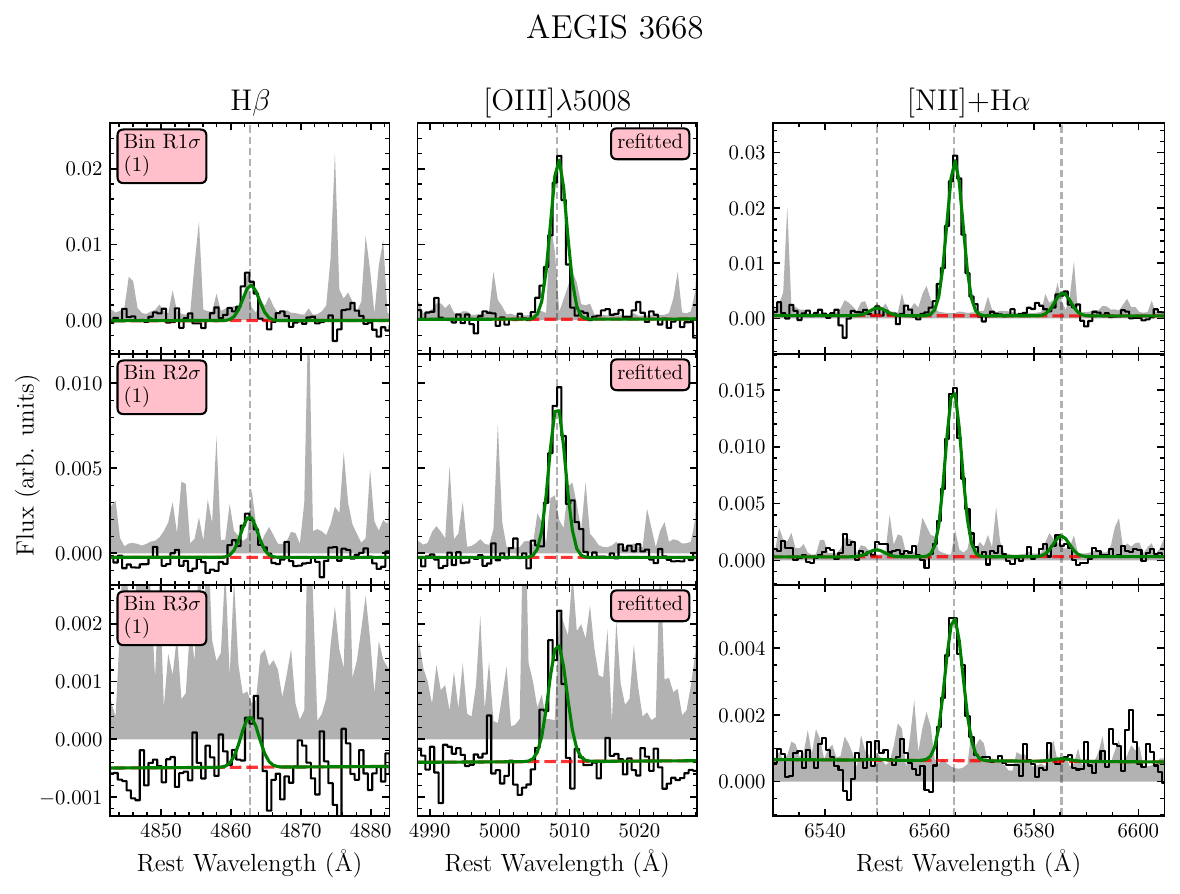} 
    \caption{Radially-binned spectra for AEGIS~3668 and fitted line profiles, similar to Fig.~\ref{fig:rad_fits/co19985_fits}.
    \label{fig:rad_fits/aeg905_fits}}
\end{figure*}

\begin{figure*}[ht!]
    \centering
    \includegraphics[width=0.8\textwidth]{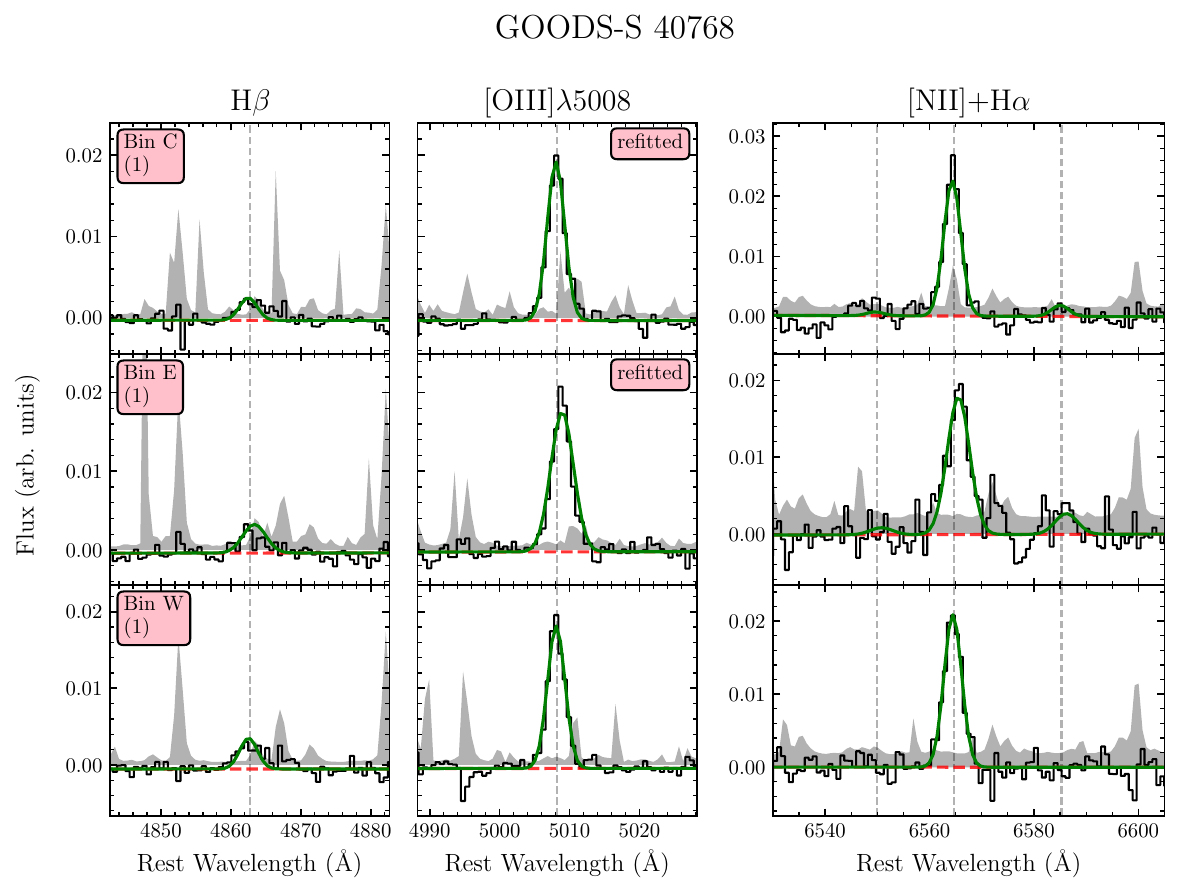} 
    \caption{Clump-binned spectra for GOODS-S~40768 and fitted line profiles, similar to Fig.~\ref{fig:rad_fits/co19985_fits}.
    \label{fig:rad_fits/gs40768_fits}}
\end{figure*}

In Figs.~\ref{fig:rad_fits/co19985_fits}--\ref{fig:rad_fits/gs40768_fits}, we showcase the emission line fitting for the radial and clump bins as defined in Section~\ref{subsubsec:radial_binning}. The color scheme follows that of Fig.~\ref{fig:int_fits}. In each row, the upper-left red text box contains the bin name and whether a one-component model (1) or two-component model (2) is preferred and plotted. A ``refitted" tag is displayed at the top right of the band that underwent refitting as determined by the routine described in Section~\ref{subsubsec:emline_fitting}. In Figs.~\ref{fig:rad_fits/co20062_fits} and \ref{fig:rad_fits/gs40768_fits}, the circular clump bins are shown. 

\subsection{Voronoi bins}\label{app_subsec:vorbin}

\begin{figure*}[ht!]
    \centering
    \includegraphics[width=0.8\textwidth]{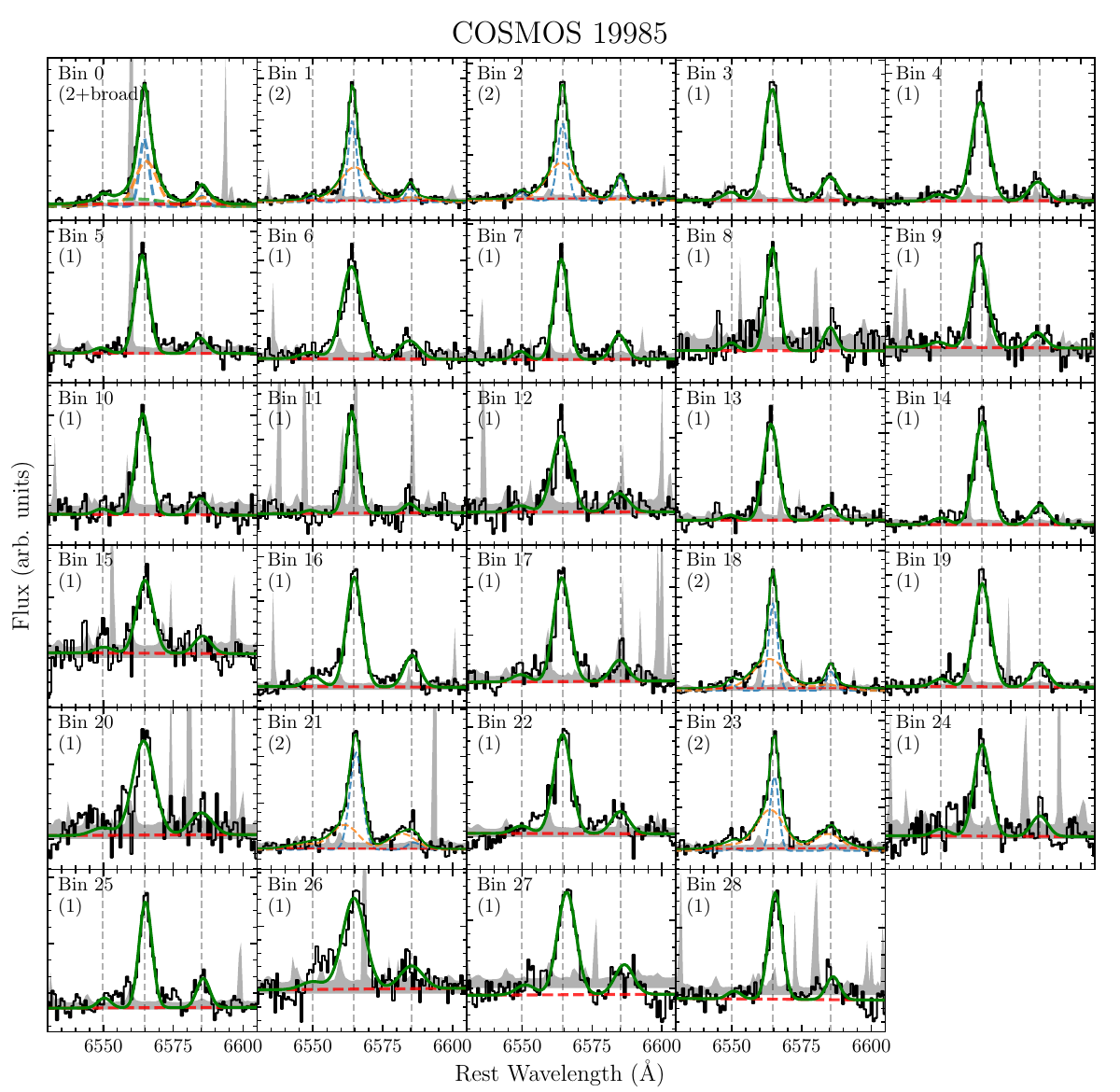} 
    \caption{Voronoi-binned K-band spectra for COSMOS~19985 and fitted line profiles. The color scheme follows that of Fig.~\ref{fig:int_fits}. The upper-left corner of each panel displays the bin name and whether a one-component model (1) or two-component model (2) is preferred and plotted.
    \label{fig:vorbin_fits/co19985_vorbin_final_fits}}
\end{figure*}

\begin{figure*}[ht!]
    \centering
    \includegraphics[width=0.8\textwidth]{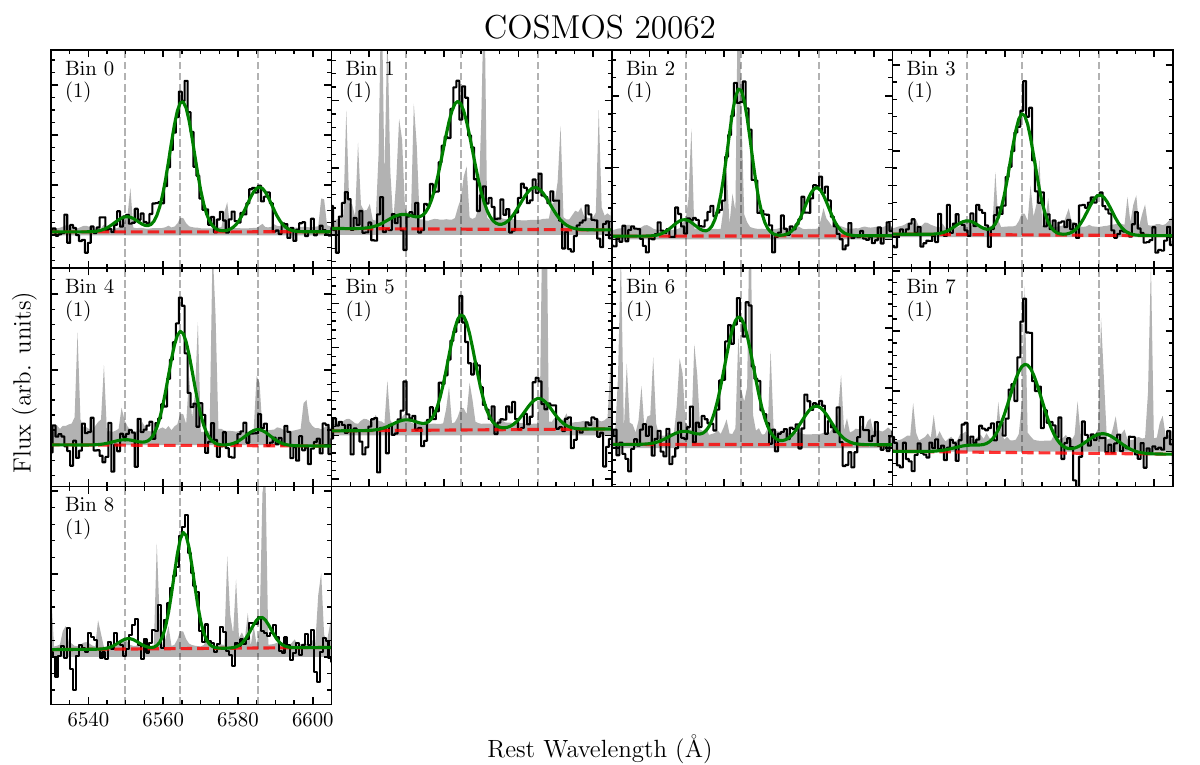} 
    \caption{Voronoi-binned K-band spectra for COSMOS~20062 and fitted line profiles, similar to Fig.~\ref{fig:rad_fits/co19985_fits}.
    \label{fig:vorbin_fits/co20062_vorbin_final_fits}}
\end{figure*}

\begin{figure*}[ht!]
    \centering
    \includegraphics[width=0.8\textwidth]{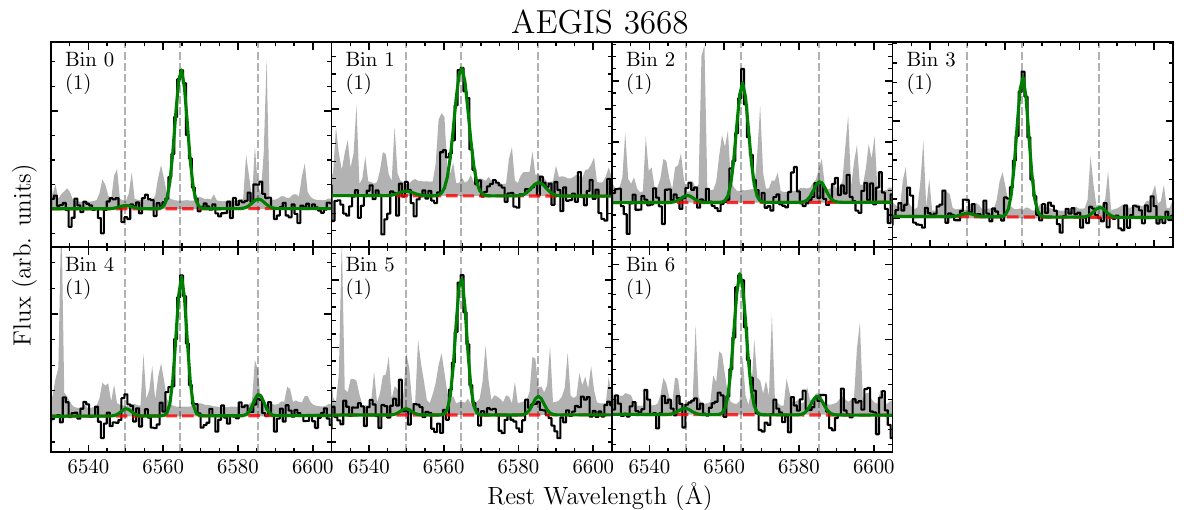} 
    \caption{Voronoi-binned K-band spectra for AEGIS~3668 and fitted line profiles, similar to Fig.~\ref{fig:rad_fits/co19985_fits}.
    \label{fig:vorbin_fits/aeg905_vorbin_final_fits}}
\end{figure*}

In Figs.~\ref{fig:vorbin_fits/co19985_vorbin_final_fits}--\ref{fig:vorbin_fits/aeg905_vorbin_final_fits}, we showcase the emission line fitting for the Voronoi bins as defined in Section~\ref{subsubsec:vorbin}. The color scheme follows that of Fig.~\ref{fig:int_fits}. The upper-left corner of each panel shows the bin name and whether a one-component model (1) or two-component model (2) is preferred and plotted. 

\section{BIC comparisons}\label{app_sec:bic}

As mentioned in Appendix Section~\ref{app_subsec:mos_int}, Fig.~\ref{fig:co20062_integrated_bic_comparison} displays spectral fits for the OSIRIS integrated spectrum of COSMOS~20062, comparing different models to assess the best-fit solution. The original one-component model, determined by the fitting routine in Section~\ref{subsubsec:emline_fitting}, is shown alongside alternative models incorporating an additional broad \halpha{} component in one- and two-component profiles, as well as a three-component model. The BIC values for each fit are displayed in the upper-right corner of each panel, with the best-fit model determined by the lowest BIC value. Despite the significant residuals, the analysis finds that the original one-component model remains the preferred fit. 

We acknowledge the discrepancy between the fits to the OSIRIS integrated spectrum and the MOSFIRE spectrum for COSMOS~20062. Although a two-component model is preferred by the higher-S/N MOSFIRE spectrum, the MOSFIRE dataset is used primarily for reference and to motivate this follow-up study. Because the analysis presented in this paper focuses on the spatially resolved information uniquely available in the OSIRIS observations, model selection is based consistently on the main OSIRIS dataset, rather than the ancillary MOSFIRE spectra.

Similarly, Fig.~\ref{fig:co19985_interesting_bic_comparison} highlights Voronoi bins in COSMOS~19985 where the initial model does not yield the lowest BIC value. Among the four bins examined, only Bin 0 (the central-most spaxel) exhibits a reduction of $\Delta\mathrm{BIC}>6$, favoring a more complex model with a narrow and broad \nii+\halpha{} component and a broad \halpha{} component.

\begin{figure*}[ht!]
    \centering
    \includegraphics[width=\textwidth]{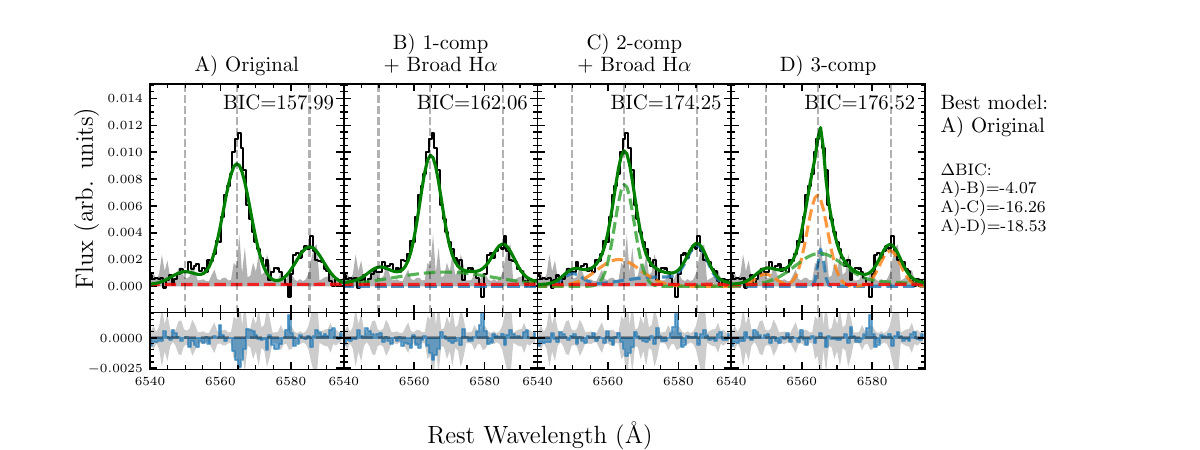} 
    \caption{Spectral fits for the OSIRIS integrated spectrum of COSMOS~20062. The color scheme follows that of Fig.~\ref{fig:int_fits}. The BIC value of each fit is quoted at the upper-right corner of each panel. Model A) is the one-component model preferred by the routine described in Section~\ref{subsubsec:emline_fitting}. Models B) and C) are one- and two-component profiles with an additional broad \halpha{} Gaussian. Model D) is a three-component model. The best model, as determined by BIC analysis, is shown in the right column, along with the differences in BIC values between models. Despite the large residuals, the original one-component model is preferred.
    \label{fig:co20062_integrated_bic_comparison}}
\end{figure*}

\begin{figure*}[ht!]
    \centering
    \includegraphics[width=\textwidth]{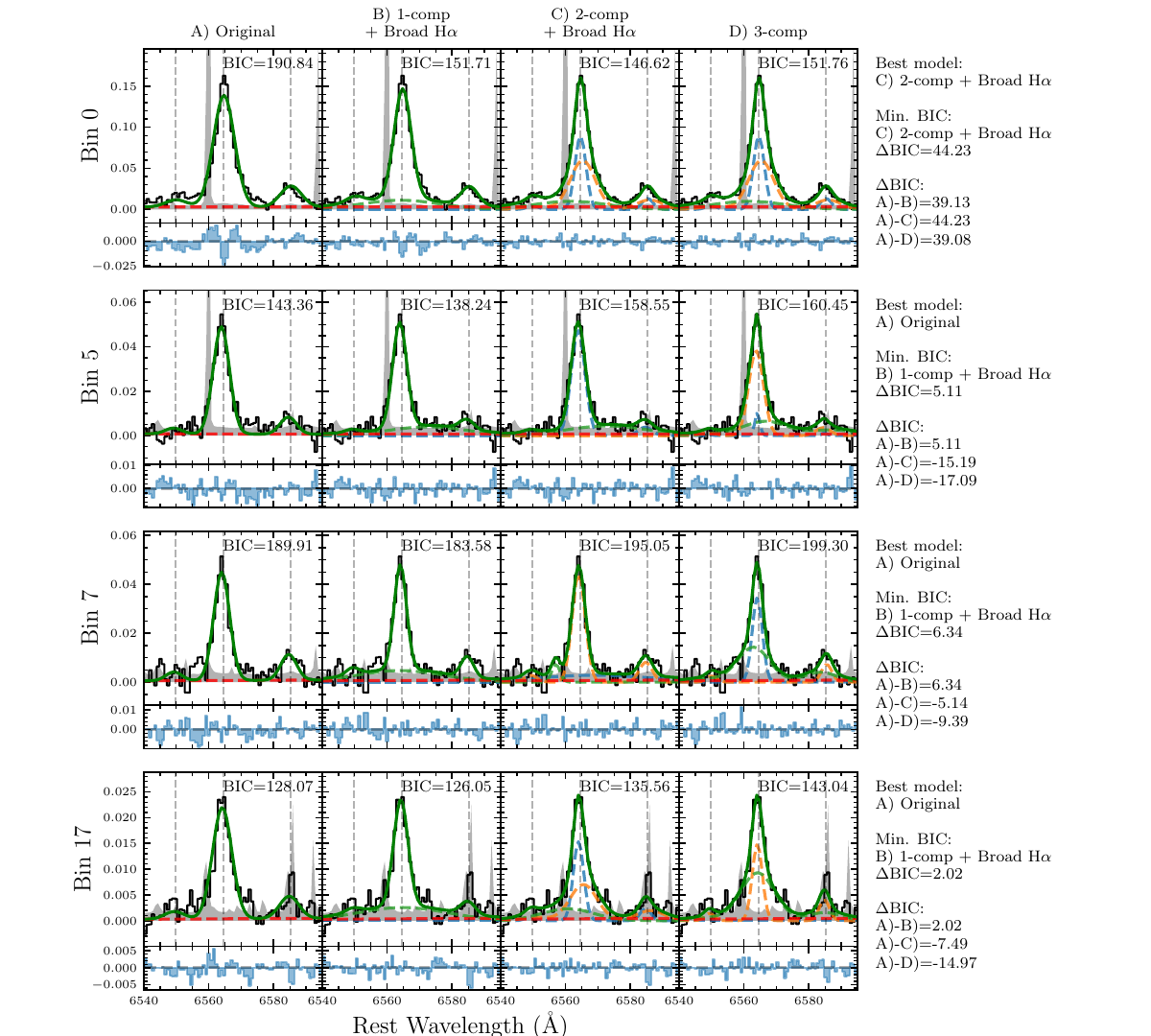} 
    \caption{Spectral fits for a sample of Voronoi bins in COSMOS~19985 where the original fit does not have the lowest BIC value. The plotting scheme is similar to that of Fig.~\ref{fig:co20062_integrated_bic_comparison}. Of the four bins shown, only Bin 0 has a reduction of $\Delta\mathrm{BIC}>6$ as compared to the original model to favor the more complex model. 
    \label{fig:co19985_interesting_bic_comparison}}
\end{figure*}

\end{document}